\documentclass[longauth,
  kazakh,
  russian,
  english]{aa}
\usepackage[LGR,T2A,T1]{fontenc}
\usepackage{CJKutf8}
\newcommand{\tinier}{\fontsize{7.5pt}{7.5pt}\selectfont}
\usepackage{setspace}

\usepackage{graphicx}
\usepackage{natbib}
\usepackage{scalerel}
\usepackage{booktabs}
\usepackage[table]{xcolor}
\newcommand*{\rt}{}%\textcolor{red}}
\newcommand*{\cswr}{\ensuremath{\{C_{\rm{SWR}}}\}}

\newcommand*{\muCGji}{\ensuremath{\mu_{\rm{C_{\it{j}}|G_{\it{i}}}}}}
\newcommand*{\muCGi}{\ensuremath{\mu_{\rm{C|G_{\it{i}}}}}}
\newcommand*{\NGri}{\ensuremath{N_{\rm{G_{\it{i}}}}}}
\newcommand*{\NCGji}{\ensuremath{N_{\rm{C_{\it{j}}|G_{\it{i}}}}}}
\newcommand*{\NCGi}{\ensuremath{N_{{\rm{C}}|G_{\it{i}}}}}
\newcommand*{\SWCt}{SW-Classify}
\newcommand*{\SWRt}{SW-Refine}
\newcommand{\ignoreme}[1]{}

\usepackage{subcaption} %For subfigures
\usepackage{txfonts}
\usepackage[pdfencoding=auto,psdextra]{hyperref}
\hypersetup{
    colorlinks=true,
    linkcolor=blue,
    filecolor=magenta,      
    urlcolor=blue,
    citecolor=blue
}
\makeatletter
\renewcommand*\aa@pageof{, page \thepage{} of \pageref*{LastPage}}
\makeatother

\usepackage[utf8]{inputenc}

 \usepackage{textcomp}

\usepackage{euclid}
\usepackage{orcidlink}

\begin{document}
%
% Put the title and authors of your (Standard Project) paper here
%

\title{Euclid Quick Data Release (Q1)} \subtitle{Refining the strong lens sample with Space Warps crowd-sourced science\thanks{This publication has been made possible by the participation of volunteers in the Space Warps project. Their contributions are individually acknowledged 
% at
% \url{https://www.zooniverse.org/projects/aprajita/space-warps-esa-euclid/about/team}.
in Appendix \ref{Ap: SW Volunteers} and on the Space Warps \href{https://www.zooniverse.org/projects/aprajita/space-warps-esa-euclid/about/team}{`The Team'} webpage.} 
}

%% please do not edit the author list once you copy it from the
%% Publication Portal -- contact ECEB Bureau for changes
\newcommand{\orcid}[1]{\orcidlink{#1}}%\href{https://orcid.org/#1}

% \newcommand{\orcid}[1]{} %% define as link to https://orcid.org/#1 if needed
% \author{\normalsize Euclid Collaboration:
%   P. Holloway$^{1}$\thanks{\email{philip.holloway@physics.ox.ac.uk}},
%     A. Verma$^{1}$, 
%     M. Walmsley,
%     P. J. Marshall,
%     A. More,
%     T. Collett, 
%     \{Neural Net leads: M. Walmsley, N. Lines, L. Leuzzi, A. Manjón-García, J. Wilde, S. Vincken, R. Pearce-Casey, I. Taufik Andika\},
%     \{SW moderators: E. Baeten, C. Macmillan\},
%     \{GJ team\},
%     \{Cutout teams: K. Rojas, T. Li, M. Walmsley, A. Melo, J. A. Acevedo Barroso, N. Lines\},
%     \{Sims teams: K. Rojas, B. Metcalf, J. A. Acevedo Barroso, A. Melo\}
%     \{Pipeline teams: B. Clement, H. Degaudenzi\},
%     \{SWLG Coordinators: G. Despali, T. Collett, S. Schuldt; F. Courbin, B. Metcalf, R. Gavazzi\},
%     \{B. Nagam, D. Sluse, C. Tortora\},
%     \{SLWG Members\},
%     order may change}
% \input{Author_List}
%%%% please do not edit the author list -- contact ECEB Bureau for changes
% \newcommand{\orcid}[1]{} %% if already defined in aa.cls: comment, or use 	
%%%% Version Monday 2nd of November 2026 04:46:55 PM UT
%%%% Assumes the new A&A style file from Oct 2025 or later
%%%% Please do not edit the author list -- contact ECEB Bureau for changes
% \newcommand{\orcid}[1]{} %% if already defined in aa.cls: comment, or use renewcommand
% \newcommand{\orcid}[1]{\,\orcidlink{#1}} %% if already defined in aa.cls: comment, or use renewcommand	

\author{Euclid Collaboration: P.~Holloway\orcid{0009-0002-8896-6100}\thanks{\email{philip.holloway@port.ac.uk}}\inst{\ref{aff1},\ref{aff2}}
\and A.~Verma\orcid{0000-0002-0730-0781}\inst{\ref{aff2}}
\and T.~E.~Collett\orcid{0000-0001-5564-3140}\inst{\ref{aff1}}
\and N.~E.~P.~Lines\orcid{0009-0004-7751-1914}\inst{\ref{aff1}}
\and T.~Li\orcid{0009-0005-5008-0381}\inst{\ref{aff1}}
\and P.~J.~Marshall\orcid{0000-0002-0113-5770}\inst{\ref{aff3},\ref{aff4}}
\and A.~More\orcid{0000-0001-7714-7076}\inst{\ref{aff5},\ref{aff6}}
\and M.~Walmsley\orcid{0000-0002-6408-4181}\inst{\ref{aff7},\ref{aff8}}
\and K.~Rojas\orcid{0000-0003-1391-6854}\inst{\ref{aff9}}
\and A.~Manj\'on-Garc\'ia\orcid{0000-0002-7413-8825}\inst{\ref{aff10}}
\and C.~Tortora\orcid{0000-0001-7958-6531}\inst{\ref{aff11}}
\and D.~Sluse\orcid{0000-0001-6116-2095}\inst{\ref{aff12}}
\and J.~A.~Acevedo~Barroso\orcid{0000-0002-9654-1711}\inst{\ref{aff13}}
\and L.~R.~Ecker\orcid{0009-0005-3508-2469}\inst{\ref{aff14},\ref{aff15}}
\and S.~Schuldt\orcid{0000-0003-2497-6334}\inst{\ref{aff16},\ref{aff17},\ref{aff18}}
\and S.~H.~Vincken\orcid{0009-0005-7305-2359}\inst{\ref{aff19}}
\and J.~Wilde\orcid{0000-0002-4460-7379}\inst{\ref{aff20}}
\and H.~J.~Dickinson\orcid{0000-0003-0475-008X}\inst{\ref{aff21}}
\and C.~Scarlata\orcid{0000-0002-9136-8876}\inst{\ref{aff22}}
\and F.~Courbin\orcid{0000-0003-0758-6510}\inst{\ref{aff20},\ref{aff23},\ref{aff24}}
\and G.~Despali\orcid{0000-0001-6150-4112}\inst{\ref{aff25},\ref{aff26},\ref{aff27}}
\and R.~B.~Metcalf\orcid{0000-0003-3167-2574}\inst{\ref{aff25},\ref{aff26}}
\and R.~Gavazzi\orcid{0000-0002-5540-6935}\inst{\ref{aff28},\ref{aff29}}
\and S.~Serjeant\orcid{0000-0002-0517-7943}\inst{\ref{aff21}}
\and K.~Jahnke\orcid{0000-0003-3804-2137}\inst{\ref{aff30}}
\and T.~T.~Thai\orcid{0000-0002-8408-4816}\inst{\ref{aff31}}
\and B.~Altieri\orcid{0000-0003-3936-0284}\inst{\ref{aff32}}
\and L.~Amendola\orcid{0000-0002-0835-233X}\inst{\ref{aff33},\ref{aff34}}
\and S.~Andreon\orcid{0000-0002-2041-8784}\inst{\ref{aff35}}
\and N.~Auricchio\orcid{0000-0003-4444-8651}\inst{\ref{aff26}}
\and C.~Baccigalupi\orcid{0000-0002-8211-1630}\inst{\ref{aff36},\ref{aff37},\ref{aff38},\ref{aff39}}
\and M.~Baldi\orcid{0000-0003-4145-1943}\inst{\ref{aff40},\ref{aff26},\ref{aff27}}
\and S.~Bardelli\orcid{0000-0002-8900-0298}\inst{\ref{aff26}}
\and P.~Battaglia\orcid{0000-0002-7337-5909}\inst{\ref{aff26}}
\and R.~Bender\orcid{0000-0001-7179-0626}\inst{\ref{aff15},\ref{aff14}}
\and A.~Biviano\orcid{0000-0002-0857-0732}\inst{\ref{aff37},\ref{aff36}}
\and M.~Bolzonella\orcid{0000-0003-3278-4607}\inst{\ref{aff26}}
\and E.~Branchini\orcid{0000-0002-0808-6908}\inst{\ref{aff41},\ref{aff42},\ref{aff35}}
\and M.~Brescia\orcid{0000-0001-9506-5680}\inst{\ref{aff43},\ref{aff11}}
\and S.~Camera\orcid{0000-0003-3399-3574}\inst{\ref{aff44},\ref{aff45},\ref{aff46}}
\and G.~Ca\~nas-Herrera\orcid{0000-0003-2796-2149}\inst{\ref{aff47}}
\and V.~Capobianco\orcid{0000-0002-3309-7692}\inst{\ref{aff46}}
\and C.~Carbone\orcid{0000-0003-0125-3563}\inst{\ref{aff18}}
\and J.~Carretero\orcid{0000-0002-3130-0204}\inst{\ref{aff48},\ref{aff49}}
\and M.~Castellano\orcid{0000-0001-9875-8263}\inst{\ref{aff50}}
\and G.~Castignani\orcid{0000-0001-6831-0687}\inst{\ref{aff26}}
\and S.~Cavuoti\orcid{0000-0002-3787-4196}\inst{\ref{aff11},\ref{aff51}}
\and A.~Cimatti\inst{\ref{aff52}}
\and C.~Colodro-Conde\inst{\ref{aff53}}
\and G.~Congedo\orcid{0000-0003-2508-0046}\inst{\ref{aff54}}
\and C.~J.~Conselice\orcid{0000-0003-1949-7638}\inst{\ref{aff8}}
\and L.~Conversi\orcid{0000-0002-6710-8476}\inst{\ref{aff55},\ref{aff32}}
\and Y.~Copin\orcid{0000-0002-5317-7518}\inst{\ref{aff56}}
\and H.~M.~Courtois\orcid{0000-0003-0509-1776}\inst{\ref{aff57}}
\and M.~Cropper\orcid{0000-0003-4571-9468}\inst{\ref{aff58}}
\and H.~Degaudenzi\orcid{0000-0002-5887-6799}\inst{\ref{aff59}}
\and G.~De~Lucia\orcid{0000-0002-6220-9104}\inst{\ref{aff37}}
\and H.~Dole\orcid{0000-0002-9767-3839}\inst{\ref{aff60}}
\and F.~Dubath\orcid{0000-0002-6533-2810}\inst{\ref{aff59}}
\and X.~Dupac\inst{\ref{aff32}}
\and S.~Dusini\orcid{0000-0002-1128-0664}\inst{\ref{aff61}}
\and S.~Escoffier\orcid{0000-0002-2847-7498}\inst{\ref{aff62}}
\and M.~Fabricius\orcid{0000-0002-7025-6058}\inst{\ref{aff15},\ref{aff14}}
\and M.~Farina\orcid{0000-0002-3089-7846}\inst{\ref{aff63}}
\and R.~Farinelli\inst{\ref{aff26}}
\and S.~Ferriol\inst{\ref{aff56}}
\and F.~Finelli\orcid{0000-0002-6694-3269}\inst{\ref{aff26},\ref{aff64}}
\and S.~Fotopoulou\orcid{0000-0002-9686-254X}\inst{\ref{aff65}}
\and M.~Frailis\orcid{0000-0002-7400-2135}\inst{\ref{aff37}}
\and E.~Franceschi\orcid{0000-0002-0585-6591}\inst{\ref{aff26}}
\and M.~Fumana\orcid{0000-0001-6787-5950}\inst{\ref{aff18}}
\and S.~Galeotta\orcid{0000-0002-3748-5115}\inst{\ref{aff37}}
\and K.~George\orcid{0000-0002-1734-8455}\inst{\ref{aff66}}
\and B.~Gillis\orcid{0000-0002-4478-1270}\inst{\ref{aff54}}
\and C.~Giocoli\orcid{0000-0002-9590-7961}\inst{\ref{aff26},\ref{aff27}}
\and J.~Gracia-Carpio\orcid{0000-0003-4689-3134}\inst{\ref{aff15}}
\and A.~Grazian\orcid{0000-0002-5688-0663}\inst{\ref{aff67}}
\and F.~Grupp\inst{\ref{aff15},\ref{aff14}}
\and L.~Guzzo\orcid{0000-0001-8264-5192}\inst{\ref{aff68},\ref{aff35},\ref{aff69}}
\and W.~G.~Hartley\orcid{0000-0002-4563-5101}\inst{\ref{aff59}}
\and S.~V.~H.~Haugan\orcid{0000-0001-9648-7260}\inst{\ref{aff70}}
\and W.~Holmes\orcid{0009-0007-8554-4646}\inst{\ref{aff13}}
\and I.~M.~Hook\orcid{0000-0002-2960-978X}\inst{\ref{aff71}}
\and F.~Hormuth\inst{\ref{aff72}}
\and A.~Hornstrup\orcid{0000-0002-3363-0936}\inst{\ref{aff73},\ref{aff74}}
\and M.~Jhabvala\inst{\ref{aff75}}
\and B.~Joachimi\orcid{0000-0001-7494-1303}\inst{\ref{aff76}}
\and S.~Kermiche\orcid{0000-0002-0302-5735}\inst{\ref{aff62}}
\and A.~Kiessling\orcid{0000-0002-2590-1273}\inst{\ref{aff13}}
\and B.~Kubik\orcid{0009-0006-5823-4880}\inst{\ref{aff56}}
\and M.~K\"ummel\orcid{0000-0003-2791-2117}\inst{\ref{aff14}}
\and M.~Kunz\orcid{0000-0002-3052-7394}\inst{\ref{aff77}}
\and H.~Kurki-Suonio\orcid{0000-0002-4618-3063}\inst{\ref{aff17},\ref{aff78}}
\and A.~M.~C.~Le~Brun\orcid{0000-0002-0936-4594}\inst{\ref{aff79}}
\and S.~Ligori\orcid{0000-0003-4172-4606}\inst{\ref{aff46}}
\and P.~B.~Lilje\orcid{0000-0003-4324-7794}\inst{\ref{aff70}}
\and V.~Lindholm\orcid{0000-0003-2317-5471}\inst{\ref{aff17},\ref{aff78}}
\and I.~Lloro\orcid{0000-0001-5966-1434}\inst{\ref{aff80}}
\and M.~Magliocchetti\orcid{0000-0001-9158-4838}\inst{\ref{aff63}}
\and G.~Mainetti\orcid{0000-0003-2384-2377}\inst{\ref{aff81}}
\and O.~Mansutti\orcid{0000-0001-5758-4658}\inst{\ref{aff37}}
\and S.~Marcin\inst{\ref{aff9}}
\and O.~Marggraf\orcid{0000-0001-7242-3852}\inst{\ref{aff82}}
\and M.~Martinelli\orcid{0000-0002-6943-7732}\inst{\ref{aff50},\ref{aff83}}
\and N.~Martinet\orcid{0000-0003-2786-7790}\inst{\ref{aff28}}
\and F.~Marulli\orcid{0000-0002-8850-0303}\inst{\ref{aff25},\ref{aff26},\ref{aff27}}
\and R.~J.~Massey\orcid{0000-0002-6085-3780}\inst{\ref{aff84}}
\and N.~Mauri\orcid{0000-0001-8196-1548}\inst{\ref{aff52},\ref{aff27}}
\and E.~Medinaceli\orcid{0000-0002-4040-7783}\inst{\ref{aff26}}
\and S.~Mei\orcid{0000-0002-2849-559X}\inst{\ref{aff85},\ref{aff86}}
\and M.~Meneghetti\orcid{0000-0003-1225-7084}\inst{\ref{aff26},\ref{aff27}}
\and E.~Merlin\orcid{0000-0001-6870-8900}\inst{\ref{aff67}}
\and G.~Meylan\orcid{0000-0001-6503-0209}\inst{\ref{aff87}}
\and P.~Monaco\orcid{0000-0003-2083-7564}\inst{\ref{aff88},\ref{aff37},\ref{aff38},\ref{aff36}}
\and A.~Mora\orcid{0000-0002-1922-8529}\inst{\ref{aff89}}
\and M.~Moresco\orcid{0000-0002-7616-7136}\inst{\ref{aff25},\ref{aff26}}
\and C.~Moretti\orcid{0000-0003-3314-8936}\inst{\ref{aff37},\ref{aff36},\ref{aff38}}
\and L.~Moscardini\orcid{0000-0002-3473-6716}\inst{\ref{aff25},\ref{aff26},\ref{aff27}}
\and C.~Neissner\orcid{0000-0001-8524-4968}\inst{\ref{aff90},\ref{aff49}}
\and R.~C.~Nichol\orcid{0000-0003-0939-6518}\inst{\ref{aff91}}
\and S.-M.~Niemi\orcid{0009-0005-0247-0086}\inst{\ref{aff92}}
\and J.~W.~Nightingale\orcid{0000-0002-8987-7401}\inst{\ref{aff93}}
\and C.~Padilla\orcid{0000-0001-7951-0166}\inst{\ref{aff90}}
\and S.~Paltani\orcid{0000-0002-8108-9179}\inst{\ref{aff59}}
\and F.~Pasian\orcid{0000-0002-4869-3227}\inst{\ref{aff37}}
\and K.~Pedersen\inst{\ref{aff94}}
\and V.~Pettorino\orcid{0000-0002-4203-9320}\inst{\ref{aff92}}
\and A.~Pezzotta\orcid{0000-0003-0726-2268}\inst{\ref{aff35}}
\and S.~Pires\orcid{0000-0002-0249-2104}\inst{\ref{aff95}}
\and G.~Polenta\orcid{0000-0003-4067-9196}\inst{\ref{aff96}}
\and M.~Poncet\inst{\ref{aff97}}
\and L.~A.~Popa\inst{\ref{aff98}}
\and F.~Raison\orcid{0000-0002-7819-6918}\inst{\ref{aff15}}
\and A.~Renzi\orcid{0000-0001-9856-1970}\inst{\ref{aff99},\ref{aff61},\ref{aff26}}
\and J.~Rhodes\orcid{0000-0002-4485-8549}\inst{\ref{aff13}}
\and G.~Riccio\inst{\ref{aff11}}
\and I.~Risso\orcid{0000-0003-2525-7761}\inst{\ref{aff41},\ref{aff42},\ref{aff35}}
\and E.~Romelli\orcid{0000-0003-3069-9222}\inst{\ref{aff37}}
\and M.~Roncarelli\orcid{0000-0001-9587-7822}\inst{\ref{aff26}}
\and B.~Rusholme\orcid{0000-0001-7648-4142}\inst{\ref{aff100}}
\and R.~Saglia\orcid{0000-0003-0378-7032}\inst{\ref{aff14},\ref{aff15}}
\and Z.~Sakr\orcid{0000-0002-4823-3757}\inst{\ref{aff101},\ref{aff102},\ref{aff103}}
\and A.~G.~S\'anchez\orcid{0000-0003-1198-831X}\inst{\ref{aff15}}
\and D.~Sapone\orcid{0000-0001-7089-4503}\inst{\ref{aff104}}
\and B.~Sartoris\orcid{0000-0003-1337-5269}\inst{\ref{aff14},\ref{aff37}}
\and P.~Schneider\orcid{0000-0001-8561-2679}\inst{\ref{aff82}}
\and A.~Secroun\orcid{0000-0003-0505-3710}\inst{\ref{aff62}}
\and P.~Simon\inst{\ref{aff82}}
\and C.~Sirignano\orcid{0000-0002-0995-7146}\inst{\ref{aff99},\ref{aff61}}
\and G.~Sirri\orcid{0000-0003-2626-2853}\inst{\ref{aff27}}
\and L.~Stanco\orcid{0000-0002-9706-5104}\inst{\ref{aff61}}
\and P.~Tallada-Cresp\'{i}\orcid{0000-0002-1336-8328}\inst{\ref{aff48},\ref{aff49}}
\and A.~N.~Taylor\inst{\ref{aff54}}
\and H.~I.~Teplitz\orcid{0000-0002-7064-5424}\inst{\ref{aff105}}
\and I.~Tereno\orcid{0000-0002-4537-6218}\inst{\ref{aff106},\ref{aff107}}
\and S.~Toft\orcid{0000-0003-3631-7176}\inst{\ref{aff108},\ref{aff109}}
\and R.~Toledo-Moreo\orcid{0000-0002-2997-4859}\inst{\ref{aff110},\ref{aff111}}
\and F.~Torradeflot\orcid{0000-0003-1160-1517}\inst{\ref{aff49},\ref{aff48}}
\and I.~Tutusaus\orcid{0000-0002-3199-0399}\inst{\ref{aff112},\ref{aff113},\ref{aff102}}
\and J.~Valiviita\orcid{0000-0001-6225-3693}\inst{\ref{aff17},\ref{aff78}}
\and T.~Vassallo\orcid{0000-0001-6512-6358}\inst{\ref{aff37},\ref{aff66}}
\and A.~Veropalumbo\orcid{0000-0003-2387-1194}\inst{\ref{aff35},\ref{aff42},\ref{aff41}}
\and Y.~Wang\orcid{0000-0002-4749-2984}\inst{\ref{aff100}}
\and J.~Weller\orcid{0000-0002-8282-2010}\inst{\ref{aff14},\ref{aff15}}
\and A.~Zacchei\orcid{0000-0003-0396-1192}\inst{\ref{aff37},\ref{aff36}}
\and G.~Zamorani\orcid{0000-0002-2318-301X}\inst{\ref{aff26}}
\and F.~M.~Zerbi\orcid{0000-0002-9996-973X}\inst{\ref{aff35}}
\and E.~Zucca\orcid{0000-0002-5845-8132}\inst{\ref{aff26}}
\and M.~Ballardini\orcid{0000-0003-4481-3559}\inst{\ref{aff114},\ref{aff115},\ref{aff26}}
\and P.~Bergamini\orcid{0000-0003-1383-9414}\inst{\ref{aff26}}
\and S.~Borgani\orcid{0000-0001-6151-6439}\inst{\ref{aff88},\ref{aff36},\ref{aff37},\ref{aff38},\ref{aff116}}
\and E.~Bozzo\orcid{0000-0002-8201-1525}\inst{\ref{aff59}}
\and C.~Burigana\orcid{0000-0002-3005-5796}\inst{\ref{aff117},\ref{aff64}}
\and R.~Cabanac\orcid{0000-0001-6679-2600}\inst{\ref{aff102}}
\and A.~Cappi\inst{\ref{aff118},\ref{aff26}}
\and T.~Castro\orcid{0000-0002-6292-3228}\inst{\ref{aff37},\ref{aff38},\ref{aff36},\ref{aff116}}
\and J.~A.~Escartin~Vigo\inst{\ref{aff15}}
\and J.~Garc\'ia-Bellido\orcid{0000-0002-9370-8360}\inst{\ref{aff101}}
\and T.~Gasparetto\orcid{0000-0002-7913-4866}\inst{\ref{aff50}}
\and A.~Loureiro\orcid{0000-0002-4371-0876}\inst{\ref{aff119},\ref{aff120}}
\and J.~Macias-Perez\orcid{0000-0002-5385-2763}\inst{\ref{aff121}}
\and R.~Maoli\orcid{0000-0002-6065-3025}\inst{\ref{aff122},\ref{aff50}}
\and M.~P\"ontinen\orcid{0000-0001-5442-2530}\inst{\ref{aff17}}
\and E.~Sarpa\orcid{0000-0002-1256-655X}\inst{\ref{aff37}}
\and V.~Scottez\orcid{0009-0008-3864-940X}\inst{\ref{aff123},\ref{aff124}}
\and M.~Sereno\orcid{0000-0003-0302-0325}\inst{\ref{aff26},\ref{aff27}}
\and M.~Tenti\orcid{0000-0002-4254-5901}\inst{\ref{aff27}}
\and M.~Tucci\inst{\ref{aff59}}
\and M.~Wiesmann\orcid{0009-0000-8199-5860}\inst{\ref{aff70}}
\and Y.~Akrami\orcid{0000-0002-2407-7956}\inst{\ref{aff101},\ref{aff125}}
\and I.~T.~Andika\orcid{0000-0001-6102-9526}\inst{\ref{aff14}}
\and G.~Angora\orcid{0000-0002-0316-6562}\inst{\ref{aff11},\ref{aff114}}
\and S.~Anselmi\orcid{0000-0002-3579-9583}\inst{\ref{aff61},\ref{aff99},\ref{aff126}}
\and M.~Archidiacono\orcid{0000-0003-4952-9012}\inst{\ref{aff68},\ref{aff69}}
\and G.~Arico\orcid{0000-0002-2802-2928}\inst{\ref{aff27}}
\and F.~Atrio-Barandela\orcid{0000-0002-2130-2513}\inst{\ref{aff127}}
\and M.~Baes\orcid{0000-0002-3930-2757}\inst{\ref{aff128}}
\and L.~Bazzanini\orcid{0000-0003-0727-0137}\inst{\ref{aff114},\ref{aff26}}
\and D.~Bertacca\orcid{0000-0002-2490-7139}\inst{\ref{aff99},\ref{aff67},\ref{aff61}}
\and M.~Bethermin\orcid{0000-0002-3915-2015}\inst{\ref{aff129}}
\and F.~Beutler\orcid{0000-0003-0467-5438}\inst{\ref{aff54}}
\and A.~Blanchard\orcid{0000-0001-8555-9003}\inst{\ref{aff102}}
\and L.~Blot\orcid{0000-0002-9622-7167}\inst{\ref{aff130},\ref{aff79}}
\and M.~L.~Brown\orcid{0000-0002-0370-8077}\inst{\ref{aff8}}
\and S.~Bruton\orcid{0000-0002-6503-5218}\inst{\ref{aff131}}
\and A.~Calabro\orcid{0000-0003-2536-1614}\inst{\ref{aff50}}
\and B.~Camacho~Quevedo\orcid{0000-0002-8789-4232}\inst{\ref{aff36},\ref{aff39},\ref{aff37}}
\and F.~Caro\orcid{0009-0003-1053-0507}\inst{\ref{aff50}}
\and C.~S.~Carvalho\inst{\ref{aff107}}
\and F.~Cogato\orcid{0000-0003-4632-6113}\inst{\ref{aff25},\ref{aff26}}
\and S.~Conseil\orcid{0000-0002-3657-4191}\inst{\ref{aff56}}
\and A.~R.~Cooray\orcid{0000-0002-3892-0190}\inst{\ref{aff132}}
\and P.~Corcho-Caballero\orcid{0000-0001-6327-7080}\inst{\ref{aff133}}
\and B.~Csizi\orcid{0000-0003-3227-6581}\inst{\ref{aff134}}
\and O.~Cucciati\orcid{0000-0002-9336-7551}\inst{\ref{aff26}}
\and H.~Dannerbauer\orcid{0000-0001-7147-3575}\inst{\ref{aff53},\ref{aff135}}
\and T.~de~Boer\orcid{0000-0001-5486-2747}\inst{\ref{aff136}}
\and F.~De~Paolis\orcid{0000-0001-6460-7563}\inst{\ref{aff137},\ref{aff138},\ref{aff139}}
\and G.~Desprez\orcid{0000-0001-8325-1742}\inst{\ref{aff133}}
\and A.~D\'iaz-S\'anchez\orcid{0000-0003-0748-4768}\inst{\ref{aff10}}
\and S.~Di~Domizio\orcid{0000-0003-2863-5895}\inst{\ref{aff41},\ref{aff42}}
\and J.~M.~Diego\orcid{0000-0001-9065-3926}\inst{\ref{aff140}}
\and V.~Duret\orcid{0009-0009-0383-4960}\inst{\ref{aff62}}
\and M.~Y.~Elkhashab\orcid{0000-0001-9306-2603}\inst{\ref{aff37},\ref{aff38},\ref{aff88},\ref{aff36}}
\and A.~Enia\orcid{0000-0002-0200-2857}\inst{\ref{aff26}}
\and Y.~Fang\orcid{0000-0002-0334-6950}\inst{\ref{aff14}}
\and A.~Farina\orcid{0009-0000-3420-929X}\inst{\ref{aff35},\ref{aff42}}
\and A.~G.~Ferrari\orcid{0009-0005-5266-4110}\inst{\ref{aff27}}
\and A.~Finoguenov\orcid{0000-0002-4606-5403}\inst{\ref{aff17}}
\and A.~Franco\orcid{0000-0002-4761-366X}\inst{\ref{aff137},\ref{aff138},\ref{aff139}}
\and K.~Ganga\orcid{0000-0001-8159-8208}\inst{\ref{aff85}}
\and E.~Gaztanaga\orcid{0000-0001-9632-0815}\inst{\ref{aff112},\ref{aff113},\ref{aff1}}
\and Z.~Ghaffari\orcid{0000-0002-6467-8078}\inst{\ref{aff37},\ref{aff36}}
\and F.~Giacomini\orcid{0000-0002-3129-2814}\inst{\ref{aff27}}
\and F.~Gianotti\orcid{0000-0003-4666-119X}\inst{\ref{aff26}}
\and E.~J.~Gonzalez\orcid{0000-0002-0226-9893}\inst{\ref{aff112},\ref{aff141}}
\and G.~Gozaliasl\orcid{0000-0002-0236-919X}\inst{\ref{aff142},\ref{aff17}}
\and A.~Gruppuso\orcid{0000-0001-9272-5292}\inst{\ref{aff26},\ref{aff27}}
\and M.~Guidi\orcid{0000-0001-9408-1101}\inst{\ref{aff40},\ref{aff26}}
\and C.~M.~Gutierrez\orcid{0000-0001-7854-783X}\inst{\ref{aff53},\ref{aff135}}
\and A.~Hall\orcid{0000-0002-3139-8651}\inst{\ref{aff54}}
\and N.~A.~Hatch\orcid{0000-0001-5600-0534}\inst{\ref{aff143}}
\and H.~Hildebrandt\orcid{0000-0002-9814-3338}\inst{\ref{aff144}}
\and J.~Hjorth\orcid{0000-0002-4571-2306}\inst{\ref{aff94}}
\and J.~J.~E.~Kajava\orcid{0000-0002-3010-8333}\inst{\ref{aff145},\ref{aff16},\ref{aff146}}
\and Y.~Kang\orcid{0009-0000-8588-7250}\inst{\ref{aff59}}
\and V.~Kansal\orcid{0000-0002-4008-6078}\inst{\ref{aff147},\ref{aff148}}
\and D.~Karagiannis\orcid{0000-0002-4927-0816}\inst{\ref{aff114},\ref{aff149}}
\and J.~Kim\orcid{0000-0003-2776-2761}\inst{\ref{aff2}}
\and C.~C.~Kirkpatrick\inst{\ref{aff150}}
\and A.~Kov\'acs\orcid{0000-0002-5825-579X}\inst{\ref{aff151},\ref{aff152}}
\and I.~Kova{\v{c}}i{\'{c}}\orcid{0000-0001-6751-3263}\inst{\ref{aff128}}
\and K.~Koyama\orcid{0000-0001-6727-6915}\inst{\ref{aff1}}
\and S.~Kruk\orcid{0000-0001-8010-8879}\inst{\ref{aff32}}
\and M.~C.~Lam\orcid{0000-0002-9347-2298}\inst{\ref{aff54}}
\and M.~Lattanzi\orcid{0000-0003-1059-2532}\inst{\ref{aff115}}
\and F.~Lepori\orcid{0009-0000-5061-7138}\inst{\ref{aff153}}
\and G.~Leroy\orcid{0009-0004-2523-4425}\inst{\ref{aff154},\ref{aff84}}
\and J.~Lesgourgues\orcid{0000-0001-7627-353X}\inst{\ref{aff155}}
\and T.~I.~Liaudat\orcid{0000-0002-9104-314X}\inst{\ref{aff156}}
\and L.~Linke\orcid{0000-0002-2622-8113}\inst{\ref{aff134}}
\and G.~Maggio\orcid{0000-0003-4020-4836}\inst{\ref{aff37}}
\and F.~Mannucci\orcid{0000-0002-4803-2381}\inst{\ref{aff157}}
\and F.~R.~Marleau\orcid{0000-0002-1442-2947}\inst{\ref{aff134}}
\and C.~J.~A.~P.~Martins\orcid{0000-0002-4886-9261}\inst{\ref{aff158},\ref{aff159}}
\and M.~Miluzio\inst{\ref{aff32},\ref{aff160}}
\and G.~Morgante\inst{\ref{aff26}}
\and S.~Nadathur\orcid{0000-0001-9070-3102}\inst{\ref{aff1}}
\and K.~Naidoo\orcid{0000-0002-9182-1802}\inst{\ref{aff1},\ref{aff30}}
\and A.~Navarro-Alsina\orcid{0000-0002-3173-2592}\inst{\ref{aff82}}
\and A.~Nersesian\orcid{0000-0001-6843-409X}\inst{\ref{aff12}}
\and S.~Nesseris\orcid{0000-0002-0567-0324}\inst{\ref{aff101}}
\and L.~Nicastro\orcid{0000-0001-8534-6788}\inst{\ref{aff26}}
\and F.~Oppizzi\orcid{0000-0003-3904-8370}\inst{\ref{aff42}}
\and F.~Pace\orcid{0000-0001-8039-0480}\inst{\ref{aff44},\ref{aff45},\ref{aff46}}
\and D.~Paoletti\orcid{0000-0003-4761-6147}\inst{\ref{aff26},\ref{aff64}}
\and F.~Passalacqua\orcid{0000-0002-8606-4093}\inst{\ref{aff99},\ref{aff61}}
\and K.~Paterson\orcid{0000-0001-8340-3486}\inst{\ref{aff30}}
\and L.~Patrizii\inst{\ref{aff27}}
\and C.~Pattison\orcid{0000-0003-3272-2617}\inst{\ref{aff1}}
\and A.~Pisani\orcid{0000-0002-6146-4437}\inst{\ref{aff62}}
\and D.~Potter\orcid{0000-0002-0757-5195}\inst{\ref{aff161}}
\and G.~W.~Pratt\inst{\ref{aff95}}
\and S.~Quai\orcid{0000-0002-0449-8163}\inst{\ref{aff25},\ref{aff26}}
\and M.~Radovich\orcid{0000-0002-3585-866X}\inst{\ref{aff67}}
\and G.~Rodighiero\orcid{0000-0002-9415-2296}\inst{\ref{aff99},\ref{aff67}}
\and W.~Roster\orcid{0000-0002-9149-6528}\inst{\ref{aff15}}
\and S.~Sacquegna\orcid{0000-0002-8433-6630}\inst{\ref{aff162}}
\and M.~Sahl\'en\orcid{0000-0003-0973-4804}\inst{\ref{aff163}}
\and D.~B.~Sanders\orcid{0000-0002-1233-9998}\inst{\ref{aff136}}
\and A.~Schneider\orcid{0000-0001-7055-8104}\inst{\ref{aff161}}
\and D.~Sciotti\orcid{0009-0008-4519-2620}\inst{\ref{aff50},\ref{aff83}}
\and L.~C.~Smith\orcid{0000-0002-3259-2771}\inst{\ref{aff164}}
\and J.~G.~Sorce\orcid{0000-0002-2307-2432}\inst{\ref{aff165},\ref{aff60}}
\and I.~Szapudi\orcid{0000-0003-2274-0301}\inst{\ref{aff136}}
\and M.~Talia\orcid{0000-0003-4352-2063}\inst{\ref{aff25},\ref{aff26}}
\and K.~Tanidis\orcid{0000-0001-9843-5130}\inst{\ref{aff166}}
\and C.~Tao\orcid{0000-0001-7961-8177}\inst{\ref{aff62}}
\and F.~Tarsitano\orcid{0000-0002-5919-0238}\inst{\ref{aff167},\ref{aff168},\ref{aff59}}
\and G.~Testera\orcid{0000-0003-2970-766X}\inst{\ref{aff42}}
\and R.~Teyssier\orcid{0000-0001-7689-0933}\inst{\ref{aff169}}
\and S.~Tosi\orcid{0000-0002-7275-9193}\inst{\ref{aff41},\ref{aff35},\ref{aff42}}
\and A.~Troja\orcid{0000-0003-0239-4595}\inst{\ref{aff37}}
\and C.~Uhlemann\orcid{0000-0001-7831-1579}\inst{\ref{aff170},\ref{aff93}}
\and C.~Valieri\inst{\ref{aff27}}
\and A.~Venhola\orcid{0000-0001-6071-4564}\inst{\ref{aff171}}
\and D.~Vergani\orcid{0000-0003-0898-2216}\inst{\ref{aff26}}
\and G.~Vernardos\orcid{0000-0001-8554-7248}\inst{\ref{aff172},\ref{aff173}}
\and G.~Verza\orcid{0000-0002-1886-8348}\thanks{Deceased}\inst{\ref{aff174},\ref{aff175}}
\and P.~Vielzeuf\orcid{0000-0003-2035-9339}\inst{\ref{aff62}}
\and E.~Vilenius\orcid{0000-0002-6184-7681}\inst{\ref{aff58}}
\and S.~Vinciguerra\orcid{0009-0005-4018-3184}\inst{\ref{aff28}}
\and M.~von~Wietersheim-Kramsta\orcid{0000-0003-4986-5091}\inst{\ref{aff84},\ref{aff154}}
\and N.~A.~Walton\orcid{0000-0003-3983-8778}\inst{\ref{aff164}}
\and L.~Wang\orcid{0000-0002-6736-9158}\inst{\ref{aff176},\ref{aff133}}
\and A.~H.~Wright\orcid{0000-0001-7363-7932}\inst{\ref{aff144}}
\and D.~Scott\orcid{0000-0002-6878-9840}\inst{\ref{aff177}}}
										   
%%%% please do not edit the affiliation list -- contact ECEB Bureau for changes
\institute{Institute of Cosmology and Gravitation, University of Portsmouth, Portsmouth PO1 3FX, UK\label{aff1}
\and
Department of Physics, University of Oxford, Keble Road, Oxford OX1 3RH, UK\label{aff2}
\and
Kavli Institute for Particle Astrophysics \& Cosmology (KIPAC), Stanford University, Stanford, CA 94305, USA\label{aff3}
\and
SLAC National Accelerator Laboratory, 2575 Sand Hill Road, Menlo Park, CA 94025, USA\label{aff4}
\and
The Inter-University Centre for Astronomy and Astrophysics, Post Bag 4, Ganeshkhind, Pune 411007, India\label{aff5}
\and
Kavli Institute for the Physics and Mathematics of the Universe (WPI), University of Tokyo, Kashiwa, Chiba 277-8583, Japan\label{aff6}
\and
David A. Dunlap Department of Astronomy \& Astrophysics, University of Toronto, 50 St George Street, Toronto, Ontario M5S 3H4, Canada\label{aff7}
\and
Jodrell Bank Centre for Astrophysics, Department of Physics and Astronomy, University of Manchester, Oxford Road, Manchester M13 9PL, UK\label{aff8}
\and
University of Applied Sciences and Arts of Northwestern Switzerland, School of Computer Science, 5210 Windisch, Switzerland\label{aff9}
\and
Departamento F\'isica Aplicada, Universidad Polit\'ecnica de Cartagena, Campus Muralla del Mar, 30202 Cartagena, Murcia, Spain\label{aff10}
\and
INAF-Osservatorio Astronomico di Capodimonte, Via Moiariello 16, 80131 Napoli, Italy\label{aff11}
\and
STAR Institute, University of Li{\`e}ge, Quartier Agora, All\'ee du six Ao\^ut 19c, 4000 Li\`ege, Belgium\label{aff12}
\and
Jet Propulsion Laboratory, California Institute of Technology, 4800 Oak Grove Drive, Pasadena, CA, 91109, USA\label{aff13}
\and
Universit\"ats-Sternwarte M\"unchen, Fakult\"at f\"ur Physik, Ludwig-Maximilians-Universit\"at M\"unchen, Scheinerstr.~1, 81679 M\"unchen, Germany\label{aff14}
\and
Max Planck Institute for Extraterrestrial Physics, Giessenbachstr. 1, 85748 Garching, Germany\label{aff15}
\and
Finnish Centre for Astronomy with ESO (FINCA), Quantum, Vesilinnantie 5, University of Turku, 20014 Turku, Finland\label{aff16}
\and
Department of Physics, P.O. Box 64, University of Helsinki, 00014 Helsinki, Finland\label{aff17}
\and
INAF-IASF Milano, Via Alfonso Corti 12, 20133 Milano, Italy\label{aff18}
\and
University of Applied Sciences and Arts of Northwestern Switzerland, School of Engineering, 5210 Windisch, Switzerland\label{aff19}
\and
Institut de Ci\`{e}ncies del Cosmos (ICCUB), Universitat de Barcelona (IEEC-UB), Mart\'{i} i Franqu\`{e}s 1, 08028 Barcelona, Spain\label{aff20}
\and
School of Physical Sciences, The Open University, Milton Keynes, MK7 6AA, UK\label{aff21}
\and
Minnesota Institute for Astrophysics, University of Minnesota, 116 Church St SE, Minneapolis, MN 55455, USA\label{aff22}
\and
Instituci\'o Catalana de Recerca i Estudis Avan\c{c}ats (ICREA), Passeig de Llu\'{\i}s Companys 23, 08010 Barcelona, Spain\label{aff23}
\and
Institut de Ciencies de l'Espai (IEEC-CSIC), Campus UAB, Carrer de Can Magrans, s/n Cerdanyola del Vall\'es, 08193 Barcelona, Spain\label{aff24}
\and
Dipartimento di Fisica e Astronomia "Augusto Righi" - Alma Mater Studiorum Universit\`a di Bologna, via Piero Gobetti 93/2, 40129 Bologna, Italy\label{aff25}
\and
INAF-Osservatorio di Astrofisica e Scienza dello Spazio di Bologna, Via Piero Gobetti 93/3, 40129 Bologna, Italy\label{aff26}
\and
INFN-Sezione di Bologna, Viale Berti Pichat 6/2, 40127 Bologna, Italy\label{aff27}
\and
Aix-Marseille Universit\'e, CNRS, CNES, LAM, Marseille, France\label{aff28}
\and
Institut d'Astrophysique de Paris, UMR 7095, CNRS, and Sorbonne Universit\'e, 98 bis boulevard Arago, 75014 Paris, France\label{aff29}
\and
Max-Planck-Institut f\"ur Astronomie, K\"onigstuhl 17, 69117 Heidelberg, Germany\label{aff30}
\and
National Astronomical Observatory of Japan, 2-21-1 Osawa, Mitaka, Tokyo 181-8588, Japan\label{aff31}
\and
ESAC/ESA, Camino Bajo del Castillo, s/n., Urb. Villafranca del Castillo, 28692 Villanueva de la Ca\~nada, Madrid, Spain\label{aff32}
\and
Institut f\"ur Theoretische Physik, University of Heidelberg, Philosophenweg 16, 69120 Heidelberg, Germany\label{aff33}
\and
New York University Abu Dhabi, PO Box 129188, Abu Dhabi, UAE, and Center for Astrophysics and Space Science (CASS), New York University Abu Dhabi, UAE\label{aff34}
\and
INAF-Osservatorio Astronomico di Brera, Via Brera 28, 20122 Milano, Italy\label{aff35}
\and
IFPU, Institute for Fundamental Physics of the Universe, via Beirut 2, 34151 Trieste, Italy\label{aff36}
\and
INAF-Osservatorio Astronomico di Trieste, Via G. B. Tiepolo 11, 34143 Trieste, Italy\label{aff37}
\and
INFN, Sezione di Trieste, Via Valerio 2, 34127 Trieste TS, Italy\label{aff38}
\and
SISSA, International School for Advanced Studies, Via Bonomea 265, 34136 Trieste TS, Italy\label{aff39}
\and
Dipartimento di Fisica e Astronomia, Universit\`a di Bologna, Via Gobetti 93/2, 40129 Bologna, Italy\label{aff40}
\and
Dipartimento di Fisica, Universit\`a di Genova, Via Dodecaneso 33, 16146, Genova, Italy\label{aff41}
\and
INFN-Sezione di Genova, Via Dodecaneso 33, 16146, Genova, Italy\label{aff42}
\and
Department of Physics "E. Pancini", University Federico II, Via Cinthia 6, 80126, Napoli, Italy\label{aff43}
\and
Dipartimento di Fisica, Universit\`a degli Studi di Torino, Via P. Giuria 1, 10125 Torino, Italy\label{aff44}
\and
INFN-Sezione di Torino, Via P. Giuria 1, 10125 Torino, Italy\label{aff45}
\and
INAF-Osservatorio Astrofisico di Torino, Via Osservatorio 20, 10025 Pino Torinese (TO), Italy\label{aff46}
\and
Leiden Observatory, Leiden University, Einsteinweg 55, 2333 CC Leiden, The Netherlands\label{aff47}
\and
Centro de Investigaciones Energ\'eticas, Medioambientales y Tecnol\'ogicas (CIEMAT), Avenida Complutense 40, 28040 Madrid, Spain\label{aff48}
\and
Port d'Informaci\'{o} Cient\'{i}fica, Campus UAB, C. Albareda s/n, 08193 Bellaterra (Barcelona), Spain\label{aff49}
\and
INAF-Osservatorio Astronomico di Roma, Via Frascati 33, 00078 Monteporzio Catone, Italy\label{aff50}
\and
INFN -- Sezione di Napoli, Via Cinthia 6, 80126, Napoli, Italy\label{aff51}
\and
Dipartimento di Fisica e Astronomia "Augusto Righi" - Alma Mater Studiorum Universit\`a di Bologna, Viale Berti Pichat 6/2, 40127 Bologna, Italy\label{aff52}
\and
Instituto de Astrof\'{\i}sica de Canarias, E-38205 La Laguna, Tenerife, Spain\label{aff53}
\and
Institute for Astronomy, University of Edinburgh, Royal Observatory, Blackford Hill, Edinburgh EH9 3HJ, UK\label{aff54}
\and
European Space Agency/ESRIN, Largo Galileo Galilei 1, 00044 Frascati, Roma, Italy\label{aff55}
\and
Universit\'e Lyon 1, CNRS/IN2P3, IP2I Lyon, Villeurbanne, France\label{aff56}
\and
Universit\'e Lyon 1, CNRS/IN2P3, IUF, IP2I Lyon, Villeurbanne, France \label{aff57}
\and
Mullard Space Science Laboratory, University College London, Holmbury St Mary, Dorking, Surrey RH5 6NT, UK\label{aff58}
\and
Department of Astronomy, University of Geneva, ch. d'Ecogia 16, 1290 Versoix, Switzerland\label{aff59}
\and
Universit\'e Paris-Saclay, CNRS, Institut d'astrophysique spatiale, 91405, Orsay, France\label{aff60}
\and
INFN-Padova, Via Marzolo 8, 35131 Padova, Italy\label{aff61}
\and
Aix-Marseille Universit\'e, CNRS/IN2P3, CPPM, Marseille, France\label{aff62}
\and
INAF-Istituto di Astrofisica e Planetologia Spaziali, via del Fosso del Cavaliere, 100, 00100 Roma, Italy\label{aff63}
\and
INFN-Bologna, Via Irnerio 46, 40126 Bologna, Italy\label{aff64}
\and
School of Physics, HH Wills Physics Laboratory, University of Bristol, Tyndall Avenue, Bristol, BS8 1TL, UK\label{aff65}
\and
University Observatory, LMU Faculty of Physics, Scheinerstr.~1, 81679 Munich, Germany\label{aff66}
\and
INAF-Osservatorio Astronomico di Padova, Via dell'Osservatorio 5, 35122 Padova, Italy\label{aff67}
\and
Dipartimento di Fisica "Aldo Pontremoli", Universit\`a degli Studi di Milano, Via Celoria 16, 20133 Milano, Italy\label{aff68}
\and
INFN-Sezione di Milano, Via Celoria 16, 20133 Milano, Italy\label{aff69}
\and
Institute of Theoretical Astrophysics, University of Oslo, P.O. Box 1029 Blindern, 0315 Oslo, Norway\label{aff70}
\and
Department of Physics, Lancaster University, Lancaster, LA1 4YB, UK\label{aff71}
\and
Felix Hormuth Engineering, Goethestr. 17, 69181 Leimen, Germany\label{aff72}
\and
Technical University of Denmark, Elektrovej 327, 2800 Kgs. Lyngby, Denmark\label{aff73}
\and
Cosmic Dawn Center (DAWN), Denmark\label{aff74}
\and
NASA Goddard Space Flight Center, Greenbelt, MD 20771, USA\label{aff75}
\and
Department of Physics and Astronomy, University College London, Gower Street, London WC1E 6BT, UK\label{aff76}
\and
Universit\'e de Gen\`eve, D\'epartement de Physique Th\'eorique and Centre for Astroparticle Physics, 24 quai Ernest-Ansermet, CH-1211 Gen\`eve 4, Switzerland\label{aff77}
\and
Helsinki Institute of Physics, Gustaf H{\"a}llstr{\"o}min katu 2, University of Helsinki, 00014 Helsinki, Finland\label{aff78}
\and
Laboratoire d'etude de l'Univers et des phenomenes eXtremes, Observatoire de Paris, Universit\'e PSL, Sorbonne Universit\'e, CNRS, 92190 Meudon, France\label{aff79}
\and
SKAO, Jodrell Bank, Lower Withington, Macclesfield SK11 9FT, UK\label{aff80}
\and
Centre de Calcul de l'IN2P3/CNRS, 21 avenue Pierre de Coubertin 69627 Villeurbanne Cedex, France\label{aff81}
\and
Universit\"at Bonn, Argelander-Institut f\"ur Astronomie, Auf dem H\"ugel 71, 53121 Bonn, Germany\label{aff82}
\and
INFN-Sezione di Roma, Piazzale Aldo Moro, 2 - c/o Dipartimento di Fisica, Edificio G. Marconi, 00185 Roma, Italy\label{aff83}
\and
Department of Physics, Institute for Computational Cosmology, Durham University, South Road, Durham, DH1 3LE, UK\label{aff84}
\and
Universit\'e Paris Cit\'e, CNRS, Astroparticule et Cosmologie, 75013 Paris, France\label{aff85}
\and
CNRS-UCB International Research Laboratory, Centre Pierre Bin\'etruy, IRL2007, CPB-IN2P3, Berkeley, USA\label{aff86}
\and
Institute of Physics, Laboratory of Astrophysics, Ecole Polytechnique F\'ed\'erale de Lausanne (EPFL), Observatoire de Sauverny, 1290 Versoix, Switzerland\label{aff87}
\and
Dipartimento di Fisica - Sezione di Astronomia, Universit\`a di Trieste, Via Tiepolo 11, 34131 Trieste, Italy\label{aff88}
\and
Telespazio UK S.L. for European Space Agency (ESA), Camino bajo del Castillo, s/n, Urbanizacion Villafranca del Castillo, Villanueva de la Ca\~nada, 28692 Madrid, Spain\label{aff89}
\and
Institut de F\'{i}sica d'Altes Energies (IFAE), The Barcelona Institute of Science and Technology, Campus UAB, 08193 Bellaterra (Barcelona), Spain\label{aff90}
\and
School of Mathematics and Physics, University of Surrey, Guildford, Surrey, GU2 7XH, UK\label{aff91}
\and
European Space Agency/ESTEC, Keplerlaan 1, 2201 AZ Noordwijk, The Netherlands\label{aff92}
\and
School of Mathematics, Statistics and Physics, Newcastle University, Herschel Building, Newcastle-upon-Tyne, NE1 7RU, UK\label{aff93}
\and
DARK, Niels Bohr Institute, University of Copenhagen, Jagtvej 155, 2200 Copenhagen, Denmark\label{aff94}
\and
Universit\'e Paris-Saclay, Universit\'e Paris Cit\'e, CEA, CNRS, AIM, 91191, Gif-sur-Yvette, France\label{aff95}
\and
Space Science Data Center, Italian Space Agency, via del Politecnico snc, 00133 Roma, Italy\label{aff96}
\and
Centre National d'Etudes Spatiales -- Centre spatial de Toulouse, 18 avenue Edouard Belin, 31401 Toulouse Cedex 9, France\label{aff97}
\and
Institute of Space Science, Str. Atomistilor, nr. 409 M\u{a}gurele, Ilfov, 077125, Romania\label{aff98}
\and
Dipartimento di Fisica e Astronomia "G. Galilei", Universit\`a di Padova, Via Marzolo 8, 35131 Padova, Italy\label{aff99}
\and
Caltech/IPAC, 1200 E. California Blvd., Pasadena, CA 91125, USA\label{aff100}
\and
Instituto de F\'isica Te\'orica UAM-CSIC, Campus de Cantoblanco, 28049 Madrid, Spain\label{aff101}
\and
Institut de Recherche en Astrophysique et Plan\'etologie (IRAP), Universit\'e de Toulouse, CNRS, UPS, CNES, 14 Av. Edouard Belin, 31400 Toulouse, France\label{aff102}
\and
Universit\'e St Joseph; Faculty of Sciences, Beirut, Lebanon\label{aff103}
\and
Departamento de F\'isica, FCFM, Universidad de Chile, Blanco Encalada 2008, Santiago, Chile\label{aff104}
\and
Infrared Processing and Analysis Center, California Institute of Technology, Pasadena, CA 91125, USA\label{aff105}
\and
Departamento de F\'isica, Faculdade de Ci\^encias, Universidade de Lisboa, Edif\'icio C8, Campo Grande, PT1749-016 Lisboa, Portugal\label{aff106}
\and
Instituto de Astrof\'isica e Ci\^encias do Espa\c{c}o, Faculdade de Ci\^encias, Universidade de Lisboa, Tapada da Ajuda, 1349-018 Lisboa, Portugal\label{aff107}
\and
Cosmic Dawn Center (DAWN)\label{aff108}
\and
Niels Bohr Institute, University of Copenhagen, Jagtvej 128, 2200 Copenhagen, Denmark\label{aff109}
\and
Universidad Polit\'ecnica de Cartagena, Departamento de Electr\'onica y Tecnolog\'ia de Computadoras,  Plaza del Hospital 1, 30202 Cartagena, Spain\label{aff110}
\and
European University of Technology EUt+, European Union\label{aff111}
\and
Institute of Space Sciences (ICE, CSIC), Campus UAB, Carrer de Can Magrans, s/n, 08193 Barcelona, Spain\label{aff112}
\and
Institut d'Estudis Espacials de Catalunya (IEEC),  Edifici RDIT, Campus UPC, 08860 Castelldefels, Barcelona, Spain\label{aff113}
\and
Dipartimento di Fisica e Scienze della Terra, Universit\`a degli Studi di Ferrara, Via Giuseppe Saragat 1, 44122 Ferrara, Italy\label{aff114}
\and
Istituto Nazionale di Fisica Nucleare, Sezione di Ferrara, Via Giuseppe Saragat 1, 44122 Ferrara, Italy\label{aff115}
\and
ICSC - Centro Nazionale di Ricerca in High Performance Computing, Big Data e Quantum Computing, Via Magnanelli 2, Bologna, Italy\label{aff116}
\and
INAF, Istituto di Radioastronomia, Via Piero Gobetti 101, 40129 Bologna, Italy\label{aff117}
\and
Universit\'e C\^{o}te d'Azur, Observatoire de la C\^{o}te d'Azur, CNRS, Laboratoire Lagrange, Bd de l'Observatoire, CS 34229, 06304 Nice cedex 4, France\label{aff118}
\and
Oskar Klein Centre for Cosmoparticle Physics, Department of Physics, Stockholm University, Stockholm, SE-106 91, Sweden\label{aff119}
\and
Astrophysics Group, Blackett Laboratory, Imperial College London, London SW7 2AZ, UK\label{aff120}
\and
Univ. Grenoble Alpes, CNRS, Grenoble INP, LPSC-IN2P3, 53, Avenue des Martyrs, 38000 Grenoble, France\label{aff121}
\and
Dipartimento di Fisica, Sapienza Universit\`a di Roma, Piazzale Aldo Moro 2, 00185 Roma, Italy\label{aff122}
\and
Institut d'Astrophysique de Paris, 98bis Boulevard Arago, 75014, Paris, France\label{aff123}
\and
ICL, Junia, Universit\'e Catholique de Lille, LITL, 59000 Lille, France\label{aff124}
\and
CERCA/ISO, Department of Physics, Case Western Reserve University, 10900 Euclid Avenue, Cleveland, OH 44106, USA\label{aff125}
\and
Laboratoire Univers et Th\'eorie, Observatoire de Paris, Universit\'e PSL, Universit\'e Paris Cit\'e, CNRS, 92190 Meudon, France\label{aff126}
\and
Departamento de F{\'\i}sica Fundamental. Universidad de Salamanca. Plaza de la Merced s/n. 37008 Salamanca, Spain\label{aff127}
\and
Universiteit Gent, Department of Physics and Astronomy, Proeftuinstraat 86 N3, 9000 Ghent, Belgium
\label{aff128}
\and
Universit\'e de Strasbourg, CNRS, Observatoire astronomique de Strasbourg, UMR 7550, 67000 Strasbourg, France\label{aff129}
\and
Center for Data-Driven Discovery, Kavli IPMU (WPI), UTIAS, The University of Tokyo, Kashiwa, Chiba 277-8583, Japan\label{aff130}
\and
California Institute of Technology, 1200 E California Blvd, Pasadena, CA 91125, USA\label{aff131}
\and
Department of Physics \& Astronomy, University of California Irvine, Irvine CA 92697, USA\label{aff132}
\and
Kapteyn Astronomical Institute, University of Groningen, PO Box 800, 9700 AV Groningen, The Netherlands\label{aff133}
\and
Universit\"at Innsbruck, Institut f\"ur Astro- und Teilchenphysik, Technikerstr. 25/8, 6020 Innsbruck, Austria\label{aff134}
\and
Universidad de La Laguna, Dpto. Astrof\'\i sica, E-38206 La Laguna, Tenerife, Spain\label{aff135}
\and
Institute for Astronomy, University of Hawaii, 2680 Woodlawn Drive, Honolulu, HI 96822, USA\label{aff136}
\and
Department of Mathematics and Physics E. De Giorgi, University of Salento, Via per Arnesano, CP-I93, 73100, Lecce, Italy\label{aff137}
\and
INFN, Sezione di Lecce, Via per Arnesano, CP-193, 73100, Lecce, Italy\label{aff138}
\and
INAF-Sezione di Lecce, c/o Dipartimento Matematica e Fisica, Via per Arnesano, 73100, Lecce, Italy\label{aff139}
\and
Instituto de F\'isica de Cantabria, Edificio Juan Jord\'a, Avenida de los Castros, 39005 Santander, Spain\label{aff140}
\and
Instituto de Astronomia Teorica y Experimental (IATE-CONICET), Laprida 854, X5000BGR, C\'ordoba, Argentina\label{aff141}
\and
Department of Computer Science, Aalto University, PO Box 15400, Espoo, FI-00 076, Finland\label{aff142}
\and
School of Physics and Astronomy, University of Nottingham, University Park, Nottingham NG7 2RD, UK\label{aff143}
\and
Ruhr University Bochum, Faculty of Physics and Astronomy, Astronomical Institute (AIRUB), German Centre for Cosmological Lensing (GCCL), 44780 Bochum, Germany\label{aff144}
\and
Department of Physics and Astronomy, Vesilinnantie 5, University of Turku, 20014 Turku, Finland\label{aff145}
\and
Serco for European Space Agency (ESA), Camino bajo del Castillo, s/n, Urbanizacion Villafranca del Castillo, Villanueva de la Ca\~nada, 28692 Madrid, Spain\label{aff146}
\and
ARC Centre of Excellence for Dark Matter Particle Physics, Melbourne, Australia\label{aff147}
\and
Centre for Astrophysics \& Supercomputing, Swinburne University of Technology,  Hawthorn, Victoria 3122, Australia\label{aff148}
\and
Department of Physics and Astronomy, University of the Western Cape, Bellville, Cape Town, 7535, South Africa\label{aff149}
\and
Department of Physics and Helsinki Institute of Physics, Gustaf H\"allstr\"omin katu 2, University of Helsinki, 00014 Helsinki, Finland\label{aff150}
\and
MTA-CSFK Lend\"ulet Large-Scale Structure Research Group, Konkoly-Thege Mikl\'os \'ut 15-17, H-1121 Budapest, Hungary\label{aff151}
\and
Konkoly Observatory, HUN-REN CSFK, MTA Centre of Excellence, Budapest, Konkoly Thege Mikl\'os {\'u}t 15-17. H-1121, Hungary\label{aff152}
\and
Departement of Theoretical Physics, University of Geneva, Switzerland\label{aff153}
\and
Department of Physics, Centre for Extragalactic Astronomy, Durham University, South Road, Durham, DH1 3LE, UK\label{aff154}
\and
Institute for Theoretical Particle Physics and Cosmology (TTK), RWTH Aachen University, 52056 Aachen, Germany\label{aff155}
\and
IRFU, CEA, Universit\'e Paris-Saclay 91191 Gif-sur-Yvette Cedex, France\label{aff156}
\and
INAF-Osservatorio Astrofisico di Arcetri, Largo E. Fermi 5, 50125, Firenze, Italy\label{aff157}
\and
Centro de Astrof\'{\i}sica da Universidade do Porto, Rua das Estrelas, 4150-762 Porto, Portugal\label{aff158}
\and
Instituto de Astrof\'isica e Ci\^encias do Espa\c{c}o, Universidade do Porto, CAUP, Rua das Estrelas, PT4150-762 Porto, Portugal\label{aff159}
\and
HE Space for European Space Agency (ESA), Camino bajo del Castillo, s/n, Urbanizacion Villafranca del Castillo, Villanueva de la Ca\~nada, 28692 Madrid, Spain\label{aff160}
\and
Department of Astrophysics, University of Zurich, Winterthurerstrasse 190, 8057 Zurich, Switzerland\label{aff161}
\and
INAF - Osservatorio Astronomico d'Abruzzo, Via Maggini, 64100, Teramo, Italy\label{aff162}
\and
Theoretical astrophysics, Department of Physics and Astronomy, Uppsala University, Box 524, 751 20 Uppsala, Sweden\label{aff163}
\and
Institute of Astronomy, University of Cambridge, Madingley Road, Cambridge CB3 0HA, UK\label{aff164}
\and
Univ. Lille, CNRS, Centrale Lille, UMR 9189 CRIStAL, 59000 Lille, France\label{aff165}
\and
Center for Astrophysics and Cosmology, University of Nova Gorica, Nova Gorica, Slovenia\label{aff166}
\and
Kobayashi-Maskawa Institute for the Origin of Particles and the Universe, Nagoya University, Chikusa-ku, Nagoya, 464-8602, Japan\label{aff167}
\and
Institute for Particle Physics and Astrophysics, Dept. of Physics, ETH Zurich, Wolfgang-Pauli-Strasse 27, 8093 Zurich, Switzerland\label{aff168}
\and
Department of Astrophysical Sciences, Peyton Hall, Princeton University, Princeton, NJ 08544, USA\label{aff169}
\and
Fakult\"at f\"ur Physik, Universit\"at Bielefeld, Postfach 100131, 33501 Bielefeld, Germany\label{aff170}
\and
Space physics and astronomy research unit, University of Oulu, Pentti Kaiteran katu 1, FI-90014 Oulu, Finland\label{aff171}
\and
Department of Physics and Astronomy, Lehman College of the CUNY, Bronx, NY 10468, USA\label{aff172}
\and
American Museum of Natural History, Department of Astrophysics, New York, NY 10024, USA\label{aff173}
\and
International Centre for Theoretical Physics (ICTP), Strada Costiera 11, 34151 Trieste, Italy\label{aff174}
\and
Center for Computational Astrophysics, Flatiron Institute, 162 5th Avenue, 10010, New York, NY, USA\label{aff175}
\and
SRON Netherlands Institute for Space Research, Landleven 12, 9747 AD, Groningen, The Netherlands\label{aff176}
\and
Department of Physics and Astronomy, University of British Columbia, Vancouver, BC V6T 1Z1, Canada\label{aff177}}    
%
% For a Key Project paperm please use instead:
%
% \title{\Euclid\/ preparation. TBD. Your title}
%
% \author{Euclid Collaboration: F.~Author, .....}
%

% \institute{$^{1}$ Sub-department of Astrophysics, University of Oxford, Denys Wilkinson Building, Keble Road, Oxford OX1 3RH, UK\\
% $^{2}$ Center for Large-Scale Structure\\
% $^{3}$ Institute for Fundamental Research \\
% $^{4}$ Institute for Astrophysics and Cosmology\\
% }

% 
% Put your abstract here:
%
% \input{Paper_Sections/1_Abstract}
\abstract{
We demonstrate the efficacy of crowd-sourced refinement of strong gravitational lens candidate samples by citizen scientists. Binary (lens/non-lens) classification by citizen scientists has been conducted previously, and here we extend such categorisation into `definite', `probable', `possible', and `non-lenses'. This grading is a required stage ahead of any strong lens sample publication and is usually carried out by (small) teams of researchers.
We compare the responses of the volunteers to those of strong lens researchers and discuss the optimal strategy for lens classification and grading in future strong lens searches. Volunteers are found to be able to correctly identify systems of each grade, and their responses can be calibrated to be comparable to those of researchers. 
Grade posteriors, defined as the probability that a strong lens system classified by the volunteers is a certain grade, converge most rapidly for non-lens systems. We find that one researcher classification is equivalent to only \rt{1.9} volunteer classifications when grading such systems.
There is significant disagreement in the grading of intermediate-grade lens candidates amongst both volunteers and researchers. Moving towards calibrated lens probabilities for future lens searches would help to account for such subjectivity. If the Q1 lens search was rerun with volunteers performing lens grading (and corresponding calibration), a completeness of 
$79\%$ %413 A+B lenses out of 521 for SWR over SWR SLDE-only sample in the top 1k 
for Grades A+B (corresponding to definite + probable lenses respectively) could be achieved with 
$41\%$ %413 out of 1000 
purity prior to researcher inspection -- this would correspond to an expected 
$12\,500$ %420*1900/63.1
true lenses (i.e. excluding contaminants) in Euclid DR1.
}
% This is a simple template for authors to write new MNRAS papers.
% The abstract should briefly describe the aims, methods, and main results of the paper.
% It should be a single paragraph not more than 250 words (200 words for Letters).
% No references should appear in the abstract.
%
% Provide up to five key words:
%
%    from the list in
%     https://www.aanda.org/for-authors/latex-issues/information-files#pop}
%
% Add short versions of title and author list for page headings
%
   \titlerunning{Refining the Strong Lens Sample with Q1}
   \authorrunning{Euclid Collaboration: P.~Holloway et al.}
   
   \maketitle

\nolinenumbers
%
%-------------------------------------------------------------------
%
%
%   Start the main text of your paper here
%
% \input{Paper_Sections/2_Introduction}
\section{Introduction}\label{Introduction}
Strong lensing is the deflection of light by gravity, where the observer sees multiple images of the lensed light source. Such deflection is typically caused by the most massive objects in the Universe, such as massive galaxies or galaxy groups and clusters. Strongly lensed systems are rare, requiring the close angular alignment of two galaxies (the lens and the background source) on the sky in order to produce detectable distortion.

To date, thousands of strong lens candidates have been identified through a variety of space-based and ground-based telescopes and surveys, including 
\Euclid{} \citep[e.g.,][]{Q1-SP048}, 
the \emph{James Webb} Space Telescope (JWST, e.g., \citealp{Nightingale2025,Mahler2025,Hogg2025}),
the Dark Energy Spectroscopic Instrument (DESI, e.g., \citealp{Hsu2026,Karp2026,Huang2026}),
the \emph{Hubble} Space Telescope (HST, e.g., \citealp{Faure2008,Jackson2008}), 
Hyper Suprime-Cam (HSC, e.g., \citealp{Sonnenfeld2018,Canameras2021,Shu2022,Schuldt2025}), 
the Dark Energy Survey (DES, e.g., \citealp{Jacobs2019a,Jacobs2019b,Gonzalez2025}),
% the Kilo-Degree Survey (KiDS; \citealp{Petrillo2017,Petrillo2019,He2020,Li2023}) and the
% Canada–France–Hawaii Telescope Legacy Survey (CFHTLS; \citealp{Marshall2016,More2016,Jacobs2017})
the Herschel Astrophysical Terahertz Large Area Survey (H-ATLAS, \citealp{Negrello2010}), and the Cosmic Lens All-Sky Survey (CLASS, \citealp{Myers2003,Browne2003}).

The Euclid Wide Survey (EWS; \citealp{EuclidSkyOverview}) aims to survey $14\,000\,\deg^2$ of the sky over a period of five years, to a depth of \IE{}\,$\approx$\,26.2 \citep{Scaramella-EP1}, with a PSF FWHM of $\ang{;;0.16}$ in \IE{} \citep{Q1-TP002,EuclidSkyVIS}. 
This provides a significant opportunity to expand the number of known lenses. The Euclid Quick Data Release (Q1; \citealp{Q1-TP001,Q1cite}) data set provided the first wide-area \Euclid{} data at EWS depth and served as a test-bed for strong lens classifiers. The Strong Lens Discovery Engine (\citealp{Q1-SP048}; \citealp{Q1-SP052}; \citealp{Q1-SP053}; \citealp{Q1-SP054}; \citealp{Q1-SP059}; \citealp{Ecker2026}, hereafter SLDE A--F respectively) presented a systematic strong lens search across the $63.1\,\deg^2$ area of Q1. The Q1 data were taken at the nominal EWS depth, and therefore provided a representative lens sample for the EWS. The SLDE Q1 lens search was conducted in several stages; a machine-learning (ML) search with multiple classifiers (see SLDE-C) was followed by an inspection by citizen scientists. Finally, strong lens researchers inspected the highest-scoring systems (approximately $7500$) resulting from both stages in a phase termed `Galaxy Judges' (GJ), in which researchers identified a total of 497 high-grade strong lens systems in the Q1 data. SLDE-F presented an extension to this search, broadening the catalogue-level cuts to include lower redshift lenses previously excluded. This series illustrated how the combination of \Euclid{}'s spatial resolution and wide-area capability will deliver large samples of high-probability lens candidates (around 8 per square degree, totalling over $100\,000$ for the full survey), revolutionising the field of strong lensing.

The next \Euclid data release, Data Release 1 (DR1), will comprise \rt{1900$\,\deg^2$}, a \rt{$30$}-fold increase from Q1.
%in which we expect to identify around \rt{$21\,000$} lenses.\footnote{Based on finding \rt{$715$} lenses in the Q1 area from the searches of SLDE-A, SLDE-F \citep{Ecker2026}, \citet{Xu2026} and \citet{Vincken2026}} %Footnote goes after punctuation (EC style guide). 
%%
For the DR1 researcher grading stage alone, 
% more than \rt{$100\,000$} 
% $\mathcal{O}(10^5)$
of order $10^5$
researcher classifications will be required\footnote{Based on finding \rt{$715$} lenses 
% finding $21\,000$ Grade A+B lenses, 
in the Q1 area (SLDE-A; SLDE-F; \citealp{Xu2026, Vincken2026}),
requiring six inspectors for each and even neglecting any false positives that will inevitably be contained in the sample.} to achieve the minimum of six classifications per system that have been shown to be needed to mitigate subjectivity and disagreement between researchers \citep{Rojas2023}. Strong lens grading is rarely clear-cut, and even strong lens researchers can firmly disagree on whether a system is strongly lensed or not, for example, due to ambiguous arcs or faint ring galaxies appearing similar to Einstein rings. Scaling to the full DR1 will thus have a significant time cost. However, this volume of classifications is well within the limits of a citizen science project. 

The Space Warps project \citep[SW, ][]{Marshall2016,More2016} is a citizen science project dedicated to finding strong lenses (e.g., \citealp{Geach2015,Sonnenfeld2020,Gonzalez2025}). To date these searches have asked volunteers to categorise systems into `lenses' and `non-lenses', in a binary classification workflow hereafter termed \SWCt{}. The high lens detection rate of the volunteers in these searches, alongside the Q1 lens search, provided motivation to launch Space Warps Refine (\SWRt{}) whereby volunteers instead provided grades according to their confidence a given system was a lens, enabling comparison to the grades assigned by researchers.% to those of volunteers. %, when presented with lens candidates from \Euclid Q1.

The ground-truth values used throughout this work are consensus researcher grades derived from the Q1 GJ project. In the future, calibrating both researchers and volunteers using spectroscopically confirmed systems will become possible following the 4MOST Strong Lensing Spectroscopic Legacy Survey (4SLSLS, \citealp{4SLSLS2023}). The existing spectroscopic samples are too small/inhomogeneous for this, however, 4SLSLS will measure \num{10000} lens redshifts thus providing confirmation of a significant proportion of the \Euclid lens sample. Notwithstanding these future data, researcher grades are currently the best available proxy for the lensing likelihood. This work presents the results of \SWRt{} and aims to answer the following questions.
\begin{itemize}
    \item Can volunteer grades be used as a proxy for researcher grading?
    \item Can volunteer grades be combined with those of researchers, along with the results of binary lens/non-lens classification from volunteers, to generate a purer lens sample?
    \item What is the optimum strategy for forthcoming lens searches (in particular for Euclid DR1 and the EWS) accounting for the inspection budgets of both volunteers and researchers?
    
\end{itemize}

This paper is structured as follows. In Sect. \ref{Data} we discuss the images shown to the volunteers in \SWRt{}, along with the setup. In Sect. \ref{Method} we discuss the analysis methods used to process the volunteers' classifications. In Sect. \ref{Results} we present the results and discuss possible strategies for optimising the visual inspection of lens candidates in future \Euclid lens searches within the respective inspection budgets of the volunteers and researchers. We conclude in Sect. \ref{Conclusion}. 

\section{Data}\label{Data}
The Space Warps project is hosted on the Zooniverse platform,\footnote{\url{https://www.zooniverse.org/}} which is dedicated to enabling citizen-science-powered research. 
Here, volunteers are shown astronomical images and asked to classify them.  The volunteers participating in \SWRt{} were asked to grade each galaxy system according to the grading scheme
\begin{equation}
    \begin{cases}
      \text{Grade A: Definite lens,}\\
      \text{Grade B: Probable lens,}\\
      \text{Grade C: Possible lens,}\\
      \text{Grade X: Not a lens.}
    \end{cases}      
\end{equation}
A tutorial, field guide, and pedagogical pages explaining the project's science goals were provided to the volunteers to enable them to make these classifications and to learn about strong lensing science.

\begin{figure*}
\centering
\includegraphics[width=\textwidth]{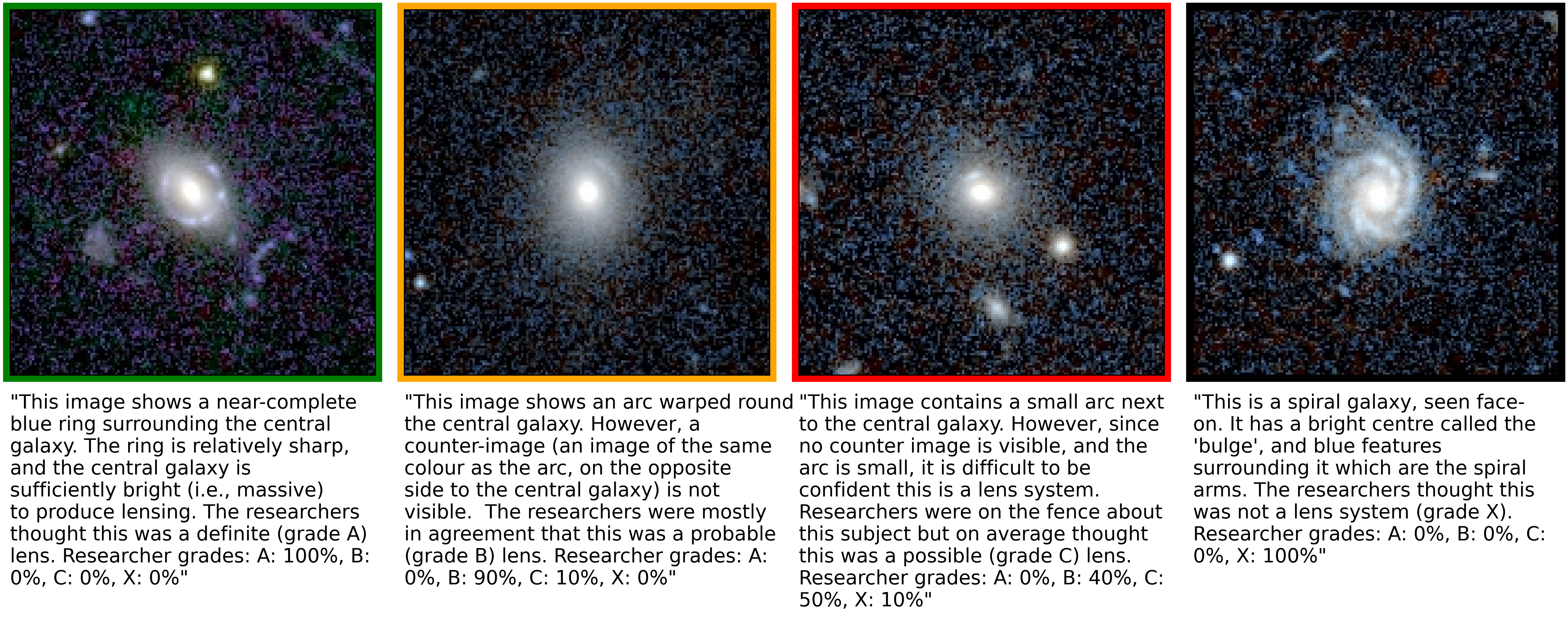}
\caption{Examples of the training systems shown to \SWRt{} volunteers, along with the feedback messages they were shown. The images are highlighted by their respective researcher grade: $\{$A: green, B: orange, C: red, X: black$\}$.}
\label{Feedback_Examples}
\end{figure*}

The volunteers participating in \SWRt{} were shown strong lens candidates that were originally shown to strong lens researchers as part of the Q1 GJ project (SLDE-A). In total, 9045 unique systems (7685 excluding simulations, see below) were shown to the volunteers.\footnote{42 systems were excluded compared to SLDE-A due to data availability and because some had been used as pedagogical examples on the SW tutorial pages, while other systems from the original GJ inspection but not identified by the systematic SLDE-A search were also included here.}
These images were split into training and test systems, though these categories were not visible to the volunteers at the point of classification and so did not influence their grade. The training systems were shown to volunteers early in their classification stream and included a feedback message specific to each system. This feedback message also indicated the grade received for that system from the GJ project. Example training systems, along with their feedback messages, are shown in Fig. \ref{Feedback_Examples}.

The \SWRt{} data set comprised 48 training systems and 8997 test systems. The test subjects classified as Grade A+B by GJ (of which there were 571), were duplicated within the Zooniverse workflow. This doubled the number of classifications received by these systems in comparison to those of a lower grade, and therefore improved the number statistics for the systems of highest interest. The test systems also included 1360 simulated lens systems (SLDE-B), which enabled us to calculate the volunteer selection function. The exact composition of the data set shown to the volunteers is described in Table \ref{T: SWR_Input_Data}.
\begin{table}
    \caption{Composition of the data set shown to the Space Warps volunteers.}
    \label{T: SWR_Input_Data}
    \centering
    \begin{tabular}{>{\raggedright\arraybackslash}p{1cm}>{\centering\arraybackslash}p{0.8cm}>{\centering\arraybackslash}p{0.8cm}>{\centering\arraybackslash}p{0.8cm}>{\centering\arraybackslash}p{0.8cm}>{\centering\arraybackslash}p{0.8cm}>{\centering\arraybackslash}p{0.9cm}}
    \hline\hline & \\[-1.5ex]
 & Gr. A& Gr. B& Gr. C& Gr. X&Sim. &Total\\\midrule
           Test&315&256&642&   6424&1360 &8997\\
           Training&10&10 &14& 
     14&0 &48\\ \midrule
 Total (unique)& 325& 266& 656& 6438& 1360&9045\\\bottomrule
\end{tabular}  
\tablefoot{The grades provided here refer to those received from the Galaxy Judges researcher-grading phase (SLDE-A), but without the weighting scheme applied in SLDE-A (Sect. \ref{S: SWAP}). The Grade A+B test systems were duplicated within the workflow, and thus received roughly twice the number of classifications from the volunteers compared to the other test systems. The volunteers were provided with feedback messages to the training images specific to each system -- these spanned all possible grades to enable the volunteers to learn key lens features to look for.}
\end{table}

In total, \rt{150\,000} classifications were received from \rt{$769$} volunteers over a period of two months. The median number of classifications received from the volunteers was \rt{$12$} -- systems were removed from the platform manually via the Panoptes-client\footnote{\url{https://github.com/zooniverse/panoptes-python-client}} once they had received more than ten classifications to ensure each system received grades from multiple volunteers. The number of training images that each volunteer had seen prior to classifying a given image varied, but their classifications were weighted according to their performance (Sect. \ref{S: SWAP}) to account for this. We present a selection of systems shown to the volunteers, alongside their \SWRt{} and GJ classifications, in Appendix \ref{Ap: Cutouts_and_Histograms}.

We compare the responses of the volunteers to those of researchers from the GJ project. The researcher crowd was composed of 
% db_GJ_class[db_GJ_class.filename.isin(db_GJ['!filename'])].user_name.unique().shape
around $50$ members of the Euclid Consortium Strong Lensing Science Working Group. These researchers made a median of 448 classifications, though with a wide spread (4--8032). There was also a wide range of responses (as expected from the results of \citealp{Rojas2023}), with some researchers much more optimistic than others (Sect. \ref{S: Performance_of_volunteers_and_researchers}). Three researchers also took part in \SWRt{}, making a total of 35 classifications. This overlap would thus have a negligible effect on the results presented here. Their skill was calculated separately for GJ and \SWRt{} classifications.

\section{Method}\label{Method}
\subsection{Space Warps Analysis Pipeline}\label{S: SWAP}
By using a subset of classified systems with a known ground truth, the volunteers' responses can be reweighted to be closer aligned to
% The large number of classifications from the volunteers enables their individual responses to be calibrated such that the output corresponds to 
a pre-set scale (e.g., lens probabilities or grades, SLDE-E, \citealp{Holloway2024}). 
Such a process is similar to that conducted in polling research, for example where certain responses are up- or down-weighted to match a given population distribution \citep{Kalton2003,Kolenikov2016}. In this work, instead of adjusting the responses to match a given population, we weight the responses according to the measured `skill' of a volunteer measured using systems with a known ground truth, thereby enabling a score to be calculated for each test system. The purpose of such calibration is to account for differences in optimism/pessimism between different volunteers when classifying whether a system is a lens or not. This process, termed the Space Warps Analysis Pipeline (SWAP), was presented in \citet{Marshall2016}, hereafter SW1. This was originally intended to calculate posterior lens probabilities, $p(\rm{Lens}|\{\it{C_j}\})$ 
% for a set of volunteers' classifications $\{C_j\}$ for a particular subject, for which non-lenses and simulated lenses were used for calibration. 
for a particular system, based on a set of volunteers' classifications $\{C_j\}$, using non-lenses and simulated lenses for calibration. 
Here, we present an extension to this procedure for processing the volunteers' grades which has four options rather than binary lens/non-lens classifications. In this case, the aim is to calibrate each volunteer's grade against the consensus grade from the researchers, for example accounting for differences in how a volunteer might assign a Grade A lens compared to the researcher crowd, since a volunteer on an individual basis  could in theory be systematically more optimistic or pessimistic than the researchers. In this work, we thus define the researcher grades as the ground truth, and calibrate the volunteers' responses accordingly.

The posterior probability that a system is grade $G_i$ (where $G_i\in\{A,B,C,X\}$), following classification $C_j$ from volunteer $u_j$ (where again, $C_j\in\{A,B,C,X\}$) is given via Bayes Theorem,
\begin{equation}\label{Eq:Grade_posterior}
    p(G_i|C_j) = \frac{p(C_j|G_i)p(G_i)}{p(C_j)}~.
\end{equation}
To be normalised, this can also be given by
\begin{equation}\label{Eq: Grade_posterior_normalised_by_sum}
    p(G_i|C_j) = \frac{p(C_j|G_i)p(G_i)}{\sum_k p(C_j|G_k)p(G_k)}~.
\end{equation}
As in SW1, we refer to $p(C_j|G_i)$ as the volunteer's skill. For simple yes/no classification this is given by a two-element vector for each volunteer. For this work, the skill is a $4\times4$ matrix, representing the probability that a volunteer will assign a grade $C_j$ to a system with a `true' grade $G_i$.  

Given the higher-dimensional skill matrix for grading, rather than yes/no classification, the matrix elements, $p(C_j|G_i)$ are more affected by statistical uncertainty -- a volunteer may have only seen one or a few `calibration' systems of each grade. Accounting for this uncertainty is important. The rate $\muCGji$ at which a volunteer classifies a grade $G_i$ system as grade $C_j$, having seen $\NGri$ systems of grade $G_i$ and having assigned $\NCGji$ of these a grade $C_j$, has the posterior $p(\muCGji|\NCGji)$.
From Bayes Theorem, we have
\begin{equation}\label{Eq: User Skill}
    p(\muCGji|\NCGji)\propto p(\muCGji)p(\NCGji|\muCGji). 
\end{equation}
We used a multinomial distribution with $\NGri$ draws for the likelihood vector $\textbf{p}(\NCGi|\muCGi)=\left(p(N_{\rm{A|G_i}}|\mu_{\rm{A|G_{\it{i}}}}),...,p(N_{\rm{X|G_{\it{i}}}}|\mu_{\rm{X|G_{\it{i}}}})\right)$. We adopted a Dirichlet distribution for the prior, $p(\muCGji)$, with all concentration values set to unity. To calculate the posterior, we took samples from the prior, weighted by the likelihood $p(\NCGji|\muCGji)$. 
The result of this sampling is a distribution of possible values of the volunteer skill matrix, $p(C_j|G_i)\equiv\muCGji$, and a corresponding distribution for the grade posterior, $p(G_i|\{C_j\})$. Unless stated otherwise, we use the \rt{median} value of this distribution as our estimator for $\muCGji$.

Throughout this work, we compare the responses of the volunteers to those of researchers. In SLDE-A, the researcher grades were calculated after assigning different weights to each researcher, rather than a simple mean average (see appendix B of SLDE-A). Here, we do not use this reweighting to enable fair comparison between \SWRt{} and GJ results. 
Instead, we assign numerical values to the grades from individual researchers, $\{\rm{A} : 3, \rm{B} : 2, \rm{C} : 1, \rm{X} : 0\}$, as in SLDE-A, and then simply calculate the mean score. As in SLDE-A, the consensus grade is then given by
\begin{equation}
    \begin{cases}
      \text{Grade A: $2\leq$ Score},\\
      \text{Grade B: $1.5\leq$ Score $<2$,}\\
      \text{Grade C: $1.0\leq$ Score $< 1.5$,}\\
      \text{Grade X: Score $<1.0$.}
    \end{cases}      
\end{equation}
These thresholds were chosen following inspection of systems close to the grade boundaries and to give broad indication as to the lens likelihood. However, they were deliberately chosen to be half-integers (rather than more specific values) to emphasise that they should not be over-interpreted.
We take this opportunity to re-emphasise that lens grading is very subjective. While in this work we calibrate the responses of the volunteers by defining the researcher grades as a ground truth, in reality researchers are not infallible and often disagree \citep{Rojas2023}. Similarly, the boundaries between grades are subjective and artificial thresholds reflect a continuum of confidence that a given system is a lens. Ideally, such calibration would be made using large data sets of spectroscopically confirmed and reliably modelled lenses (e.g., from 4SLSLS, \citealp{4SLSLS2023}) to determine the probability that a system is a lens, as discussed in \citet{Holloway2024}. Notwithstanding such future data, the ABCX grading system used here reflects what is presently widely used in the literature. It is a useful proxy for lens quality and for identifying lens systems of high scientific interest. We emphasise that throughout this paper we consider the probability that a system is a given grade, rather than the probability that a system is a lens, which would require a different ground-truth for calibration.

\section{Results and discussion}\label{Results}
\subsection{Performance of the volunteers and researchers}\label{S: Performance_of_volunteers_and_researchers}
\begin{figure*}
\centering
% %%%%%%%%%%%%%%%%%
% \begin{subfigure}{0.9\textwidth}
%   \centering
%   \includegraphics[width=\textwidth]{Images/User_Skill_ABCX_SWR_over_full_dataset_unweighted_grades.pdf}
% \caption{SWR-ABCX}
% \label{SWR_skill_ABCX}
% \end{subfigure}
% %%%%%%%%%%%%%%%%%
% \begin{subfigure}{0.9\textwidth}
% \centering
% \includegraphics[width=\textwidth]{Images/User_Skill_ABCX_GJ_over_full_dataset_unweighted_grades.pdf}
% \caption{GJ-ABCX}
% \label{GJ_skill_ABCX}
% \end{subfigure}
% %%%%%%%%%%%%%%%%%
\begin{subfigure}{0.9\textwidth}
  \centering
  \includegraphics[width=\textwidth]{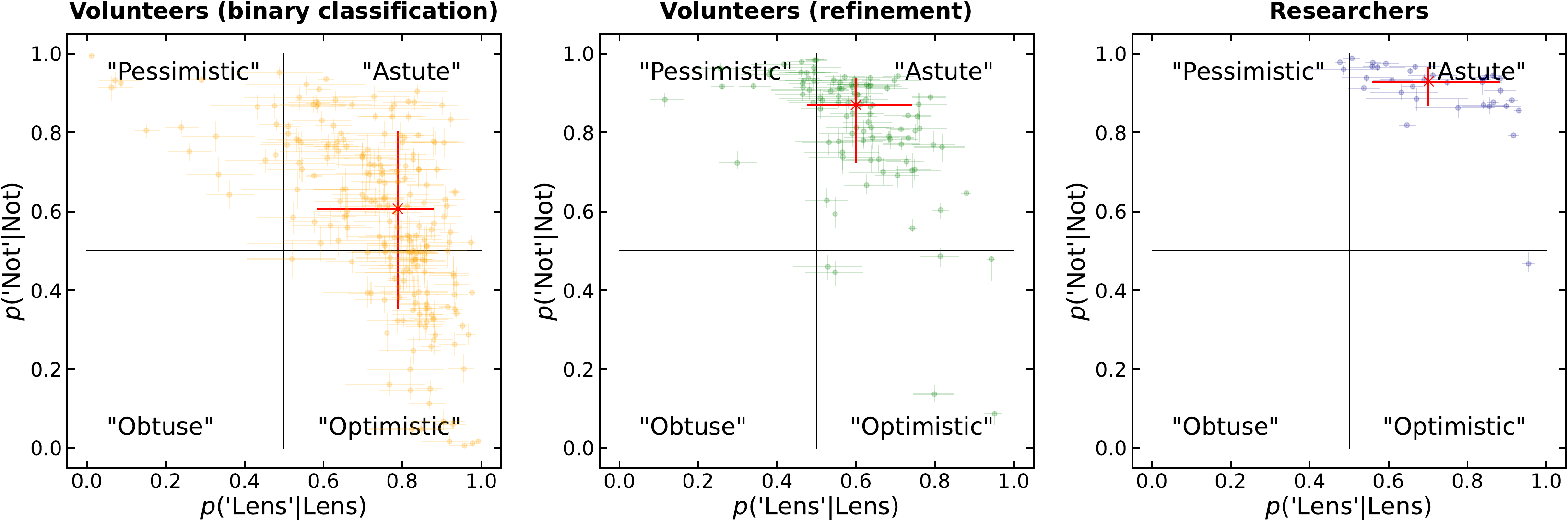}
\end{subfigure}
\caption{User skill across \SWCt{}, \SWRt{}, and GJ (left to right), respectively. We have restricted the selection to where users (either volunteers or researchers) have inspected at least \rt{100} systems with researcher grades. In the cases of \SWRt{} and GJ, we define lenses to be Grade A+B systems, and a `Lens' classification to be a classification of Grade A or Grade B. The red crosses indicate the median and $68\%$ confidence intervals on the population as a whole, while the circular scatter points indicate individual users, with error bars corresponding to the statistical uncertainty from a finite number of classifications (as defined in Sect. \ref{S: SWAP}).}
\label{Skill Plots}
\end{figure*}
%%%%%%%%%%%%%%%%%
% \begin{subfigure}{0.30\textwidth}
%   \centering
%   \includegraphics[width=\textwidth]{Images/User_Skill_GJ_over_full_dataset_unweighted_grades.pdf}
% \caption{GJ}
% \label{GJ_skill}
% \end{subfigure}
% %%%%%%%%%%%%%%%%%
% \begin{subfigure}{0.30\textwidth}
%   \centering
%   \includegraphics[width=\textwidth]{Images/User_Skill_SWC_over_full_dataset_unweighted_grades.pdf}
% \caption{SWC}
% \label{SWC_skill}
% \end{subfigure}
%%%%%%%%%%%%%%%%%
% \begin{subfigure}{0.30\textwidth}
%   \centering
%   \includegraphics[width=\textwidth]{Images/Confusion_Matrix_SWR_versus_GJ.pdf}
% \caption{SWR}
% \label{SWR_skill}
% \end{subfigure}
% %%%%%%%%%%%%%%%%%
% \begin{subfigure}{0.30\textwidth}
%   \centering
%   \includegraphics[width=\textwidth]{Images/Confusion_Matrix_GJ_versus_GJ.pdf}
% \caption{SWC}
% \label{SWC_skill}
% \end{subfigure}
% %%%%%%%%%%%%%%%%%
% \begin{subfigure}{0.30\textwidth}
%   \centering
%   \includegraphics[width=\textwidth]{Images/Confusion_Matrix_SWC_versus_GJ.pdf}
% \caption{SWC}
% \label{SWC_skill}
% \end{subfigure}
%%%%%%%%%%%%%%%%%
\begin{figure*}
\centering
\begin{subfigure}{0.9\textwidth}
  \centering
  \includegraphics[width=\textwidth]{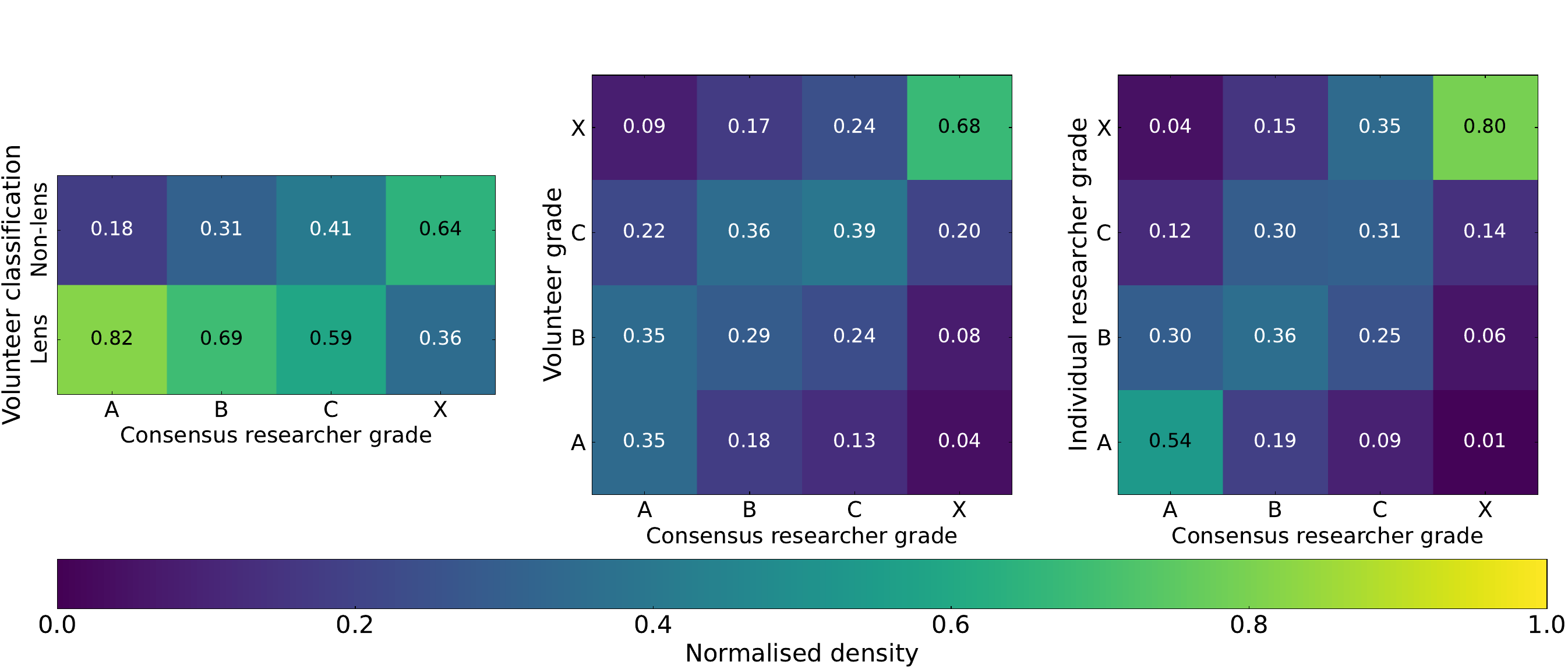}
\end{subfigure}
\caption{Confusion matrices for the user responses across \SWCt{}, \SWRt{}, and GJ (left-right). We find there is significant overlap between grades; a Grade B system is likely to be given a grade of C by at least some researchers and volunteers. The individual researchers show more agreement with the consensus researcher grade (when excluding that researchers response), and are more willing to give more definitive grades (A or X). The matrices are normalised such that the columns sum to 1.}
\label{Conf_Matrix}
\end{figure*}

%%%%%%%%%%%%%%%%%

% \label{User_Skill_Full}

In Fig. \ref{Skill Plots} we show the skill of both researchers (GJ) and volunteers. For the volunteers we show results for both \SWRt{} and \SWCt{}, the binary classification project run for Q1 (SLDE-A, SLDE-E). For ease of comparison, we treat Grade A+B systems as `Lenses' and C/X systems as `Non-lenses', defining $p(\rm{`Lens}$'$|\rm{Lens}) \equiv \it{p}\rm{(C_{\it{j}}\in\{A,B\}|G_{\it{i}}\in\{A,B}\})$ and $p(\rm{`Not}$'$|\rm{Not}) \equiv \it{p}\rm{(C_{\it{j}}\in\{C,X\}|G_{\it{i}}\in\{C,X}\})$
In all panels, the skill is evaluated against a ground truth of the consensus GJ grade.
To avoid biasing these results to favour the researchers, the skill of a given researcher is calculated by re-evaluating the consensus grade of each system using all classifications except those of that researcher. 

In all cases the majority of users reside in the top right-hand quadrant, i.e., they are `Astute', whereby their grades agree with the consensus GJ grade. However, there is large scatter in the skill of both researchers and volunteers. This likely reflects the subjective nature of lens grading, in which the boundary between, for example, a low Grade B and a high Grade C, is very blurred. By measuring each user's skill, we can account for their different biases (Eq. \ref{Eq: Grade_posterior_normalised_by_sum}) and so all users provide useful information. For \SWCt{}, the volunteers typically reside in the `Astute' or `Optimistic' quadrants. However, this is partly affected by the lens definition adopted here: we have defined Grade A+B systems as lenses, and Grade C+X systems as non-lenses. Volunteers who classified Grade C systems as lenses would be counted as optimistic, even though such systems were still sought after in the binary lens/non-lens \SWCt{} initial search.

% On average, the researchers have a higher skill than the volunteers in \SWRt{}; however, there is overlap between the two populations. A number of the volunteers have a skill equivalent or exceeding some of the researcher classifiers. This demonstrates the responses of the volunteers could be a useful proxy for those of the researchers.

Figure \ref{Conf_Matrix} shows confusion matrices for \SWCt{}, \SWRt{} and GJ responses, in comparison to the consensus (unweighted, as above) GJ grade. We find that the boundary between \SWRt{} Grade A and B candidates is blurred, reflecting a lack of confidence in classifying a system as a `definite' lens. The A-B boundary is more distinct when comparing the consensus GJ grades with the individual GJ grades from which these are derived; the researchers are more confident in their predictions than the volunteers. Finally, as expected, we find that higher grade systems are more likely to be classified as lenses (and vice versa) by \SWCt{}.

Across historical Space Warps projects (and Zooniverse projects more widely),
% including the \Euclid \SWCt{} and \SWRt{} launches,
there has been a wide distribution in the number of classifications made by volunteers; many volunteers classify just a few images, while a few classify many thousands. In \SWRt{} we find that the information content (defined as the relative entropy between the prior and posterior probabilities, averaged over all possible ground truth and classifications) continues to increase as volunteers inspect more systems; this is consistent with the finding of \citet{Marshall2016} that `Skill' correlates with `Effort'.\footnote{The information content of a particular volunteer's classifications depends on their assigned skill. This skill will evolve both as the volunteer learns (e.g. by seeing more training images) and because the SWAP skill (Eq. \ref{Eq: User Skill}) can only take fixed values (e.g. 1/2, 2/3, 3/4 etc) dependent on how many training images have been seen. Therefore the observed correlation is a combination of the effect of the volunteers' learning, and the convergence of the skill matrix.} In other words, the classifications from volunteers who have classified more calibration systems (and who will have received more training feedback) carry more information than those who made fewer. However, since the number of volunteers making fewer classifications is much larger, the entire volunteer crowd contributes valuable information.

\subsection{Grade calibration using SWAP}
% \textbf{
% \begin{itemize}
%     \item Ground truth: Published GJ grades 
%     \item Skill uncertainties: Check
% \end{itemize}
% }
\begin{figure*}
\centering
%%%%%%%%%%%%%%%%%
% \begin{subfigure}{0.9\textwidth}
%   \centering
%   \includegraphics[width=\textwidth]{Images/Raw_Grade_Bar_Chart_GJ.pdf}
% % \caption{Raw Grade}
% % \label{Raw_Grade_Bar_Chart}
% \end{subfigure}
% \begin{subfigure}{0.9\textwidth}
  \centering
  \includegraphics[width=\textwidth]{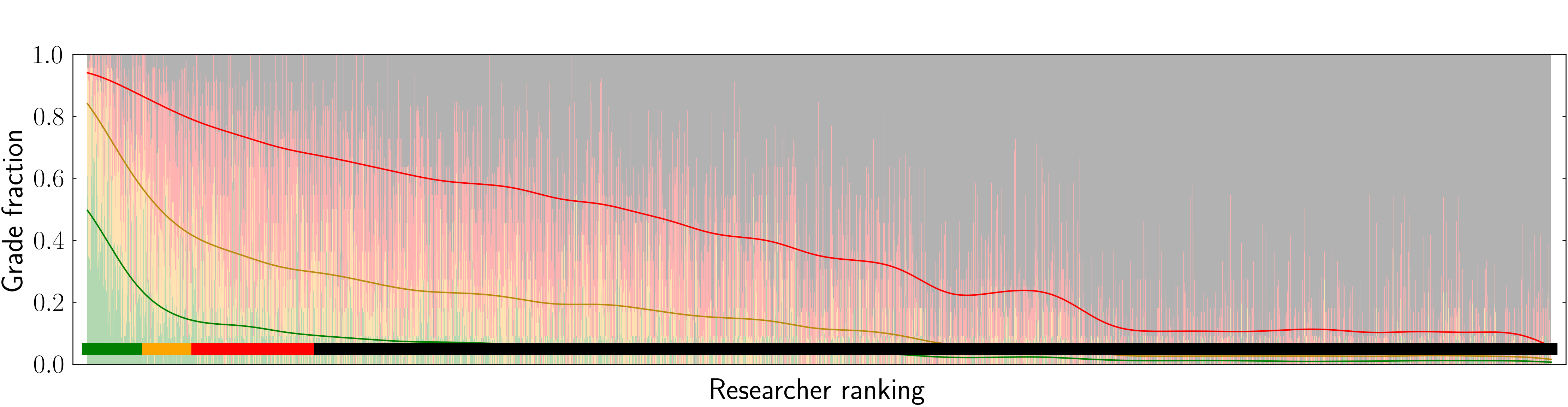}
% \caption{Raw Grade}
% \label{Raw_Grade_Bar_Chart}
% \end{subfigure}
% \begin{subfigure}{0.9\textwidth}
\centering
\includegraphics[width=\textwidth]{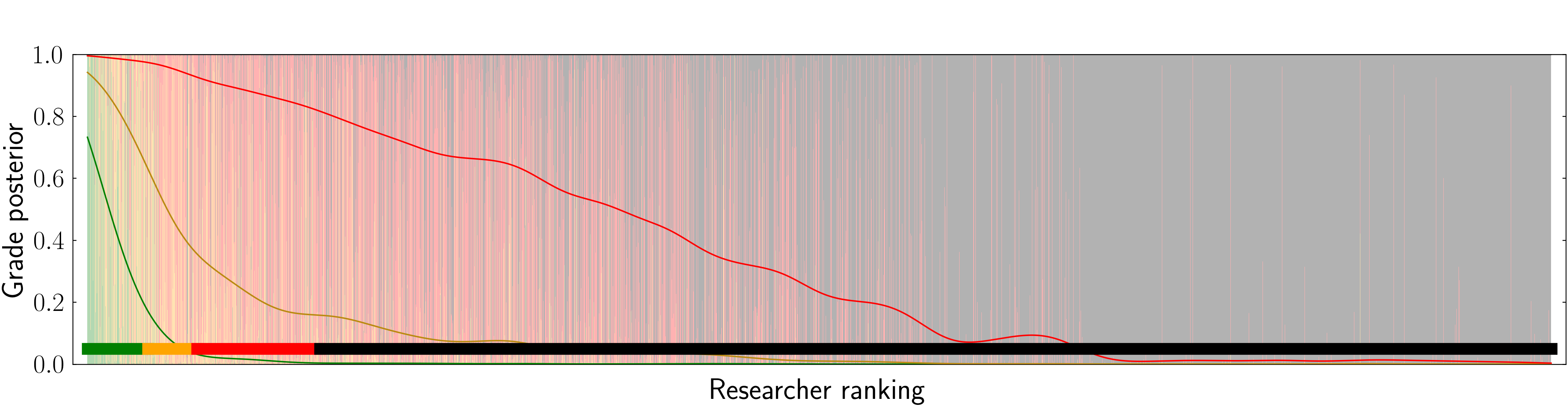}
% \caption{Posterior Grade}
% \label{Posterior_Grade_Bar_Chart}
% \end{subfigure}
\caption{\textit{Top:} Distribution of the grades assigned by the \SWRt{} volunteers to each system. Each stacked bar represents a single system, and the bars have been normalised to have a height of 1. The systems are shown in decreasing order of their GJ score ranking. The horizontal bars indicate the GJ grade boundaries, \{A: green, B: orange, C: red, X: black\}, while the solid lines show running averages. There is broad correlation between the \SWRt{} score and GJ ranking, although the volunteers are less likely to assign a Grade A than the researchers.
\textit{Bottom:} As above, but where the length of each of the stacked bars represents the grade posterior $p(G_i|\{C_j\})$ rather than the raw proportion of grades assigned for each system. The grade posteriors for a given system sum up to unity; hence, when stacked, the bars reach $y=1$. This Bayesian calibration makes the grading more confident -- the highest ranked Grade A systems are upweighted relative to the raw grade proportions, while the low ranking Grade X systems receive $p(G_i=X|\{C_j\})\approx0$ even if they received a few Grade C classifications.
% \rt{Need to rerun SWAP with unweighted grades as training subjects to make the posterior comparison valid.}
}
\label{Grade_Bar_Charts}
\end{figure*}
In Fig. \ref{Grade_Bar_Charts} we show the raw proportions of grades given to each test system by the volunteers, compared to the grade posteriors after having calibrated the volunteers responses using the subset of 48 training systems that each volunteer saw. We find that the volunteer response correlates with the GJ ranking; systems ranked higher by GJ receive higher grades from the volunteers. This is apparent even within a given grade; the highest-scoring Grade A systems from GJ receive the highest scores from \SWRt{}. The calibration from SWAP reduces some of the noise in the volunteers classifications. For example, systems that accrue predominantly `X' grades alongside a few `C' grades from the volunteers receive the grade posterior $p(X|\{C_i\})\approx1$. Similarly, systems that receive $50\%$ A-grades are up-weighted such that $p(A|\{C_i\})>0.5$. 

In summary, the calibration is useful since it accounts for the differences in behaviour of the individual graders (Fig. \ref{Skill Plots}, where some are more optimistic/pessimistic than others). This produces a purer sample of high-grade lenses than simply averaging the classifications and ensures that the grades are comparable and consistent across data sets. 
% \noindent\rt{Expand on this?}

\subsection{Optimising the inspection strategy}\label{Sect: Optimising the inspection strategy}
% \textbf{
% \begin{itemize}
%     \item Ground truth: Unweighted GJ grades.
%     \item Skill uncertainties: No. Use LOU and unweighted grades for GJ.
% \end{itemize}
% }
\begin{figure*}
\centering
\includegraphics[width=\textwidth]{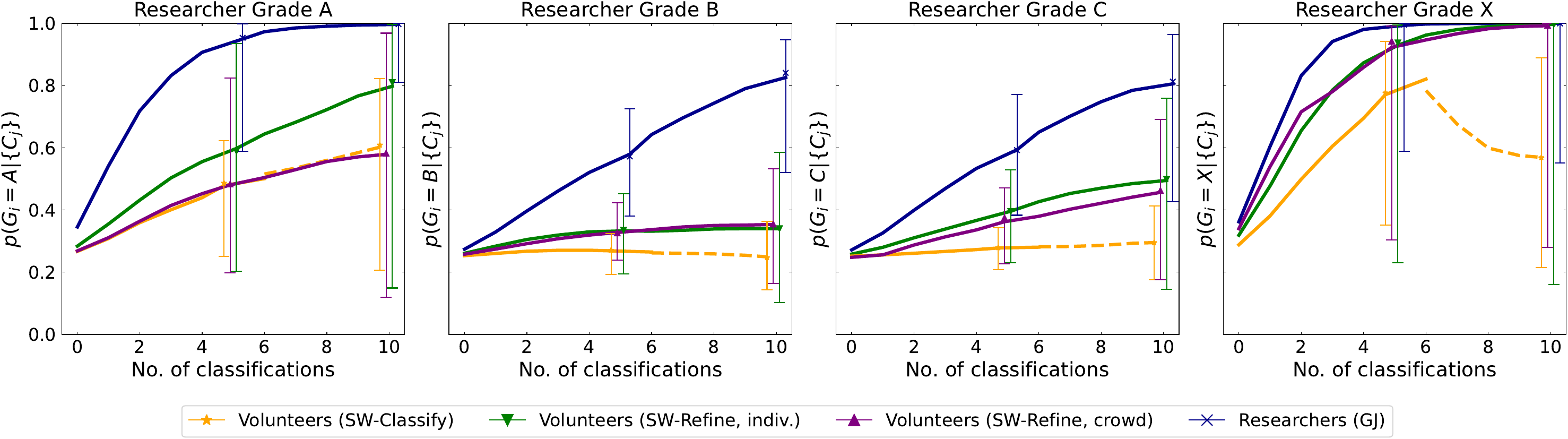}
\caption{Evolution of the grade posteriors across all available 
test systems as they receive an increasing number of classifications from \SWRt{}, \SWCt{}, and GJ members. The panels show the posteriors for systems that received researcher Grades A--X (left to right). The GJ posteriors converge the fastest, as expected since the GJ consensus grade is defined as the ground truth. \SWRt{} posteriors converge faster than those of \SWCt{}, and the fastest for Grade A and Grade X systems. The error bars include $68\%$ of the grade posteriors; their size is indicative of the subjective nature of grading.
The green curves show the grade posteriors using \SWRt{} classifications, having calibrated the volunteer responses using each individual user's skill, whereas the purple curves treat the volunteer crowd as a single super-user with a single skill matrix. In \SWCt{}, low-scoring systems were removed from the platform after six classifications, hence the selection of systems shown changes significantly beyond this point (dashed orange line). The median number of classifications per user used for calibration for \SWCt{} (orange) was 6, compared to 13 for \SWRt{} (green) and 327 for GJ (blue). 2000 classifications (500/grade) were used to calibrate the \SWRt{} super-user (purple).
% \rt{Repeat, but with LOU for the volunteers, so can then use all the grades. Similarly for the researchers?}
}
\label{Information_Content}
\end{figure*}
Given that the human inspection time of both researchers and volunteers is finite and valuable, it is important to optimise which systems are selected for human inspection. For the same reason, it is important to optimise which of the workflow(s) -- \SWCt{}, \SWRt{}, and GJ -- they are shown in, and how many classifications they receive in each. Obtaining many classifications for a system which has received many `X' or `Non-Lens' classifications would waste volunteer and/or researcher time. Figure \ref{Information_Content} shows the grade posteriors as a function of the number of classifications received from these workflows. These are split into four panels, dependent on the final researcher grade of the system. 
In the case of \SWCt{} and \SWRt{}, the volunteer's skill used to calculate the grade posterior has been computed using the full data set minus the classification for a given system, $s_i$ when calculating the grade posterior for that same system $s_i$ (hereafter the `leave-one-out' method). In the case of GJ, a given researcher's skill was calculated after recalculating the consensus researcher grades, leaving out that researcher's classifications across the whole data set.
% We show the number of classifications from SWC and \SWRt{} required to give a equivalent shift in the grade posterior as a single classification from GJ. 
Since we are using the final GJ grade as ground truth, the GJ grade posteriors can be considered upper-limits as to their information content with respect to the other classification methods.

We find that the grade posteriors converge the fastest (i.e., with the fewest number of classifications) for Grade A and X systems. It is likely that these are the easiest to classify, since they should be unambiguous examples of lenses or non-lenses. For systems of all grades, the convergence is fastest for GJ, followed by \SWRt{} and finally \SWCt{}. For Grade X systems, we find one classification from a researcher provides an equivalent posterior to 1.9 classifications from a \SWRt{} volunteer; researchers have similar performance to volunteers for identifying non-lenses. Removing \SWRt{} Grade X systems from the researcher-grading data set would reduce the time burden of researcher grading. The \SWRt{} convergence is slower for Grade A lenses than for Grade X lenses, with a larger difference with respect to the researchers; however, given enough volunteer responses (at least $10$) we can still reach grade posteriors $p(G_i=A|\{C_{\rm{SWR}}\})\approx0.8$. The convergence for Grade A systems becomes faster when combining both A and B grade posteriors, $p(G_i\in\{A,B\}|\cswr)$, since some Grade A lenses are classified by the volunteers as Grade B. Convergence is slowest for \SWCt{} across all grades. However, since this only uses a lens/non-lens classification, whereas \SWRt{} used an A/B/C/X grading scheme, it is unsurprising that the latter can provide more information. \SWCt{} has been optimised for classification speed, to remove the vast majority of galaxies that pass initial selection cuts (e.g., in magnitude) but are not lensed. A staged approach of conducting a binary classification followed by a refinement stage helps to reduce the number of systems that need the more time-intensive grading step.

The typical scatter in the posterior medians in Fig. \ref{Information_Content} is large, and larger than the typical posterior width. This scatter is due to the subjective nature of the grades which can cause the posterior for a given system to converge to an adjacent grade to that of the researchers. As is discussed in Sect. \ref{S: Extrapolating to DR1/EWS}, a purer sample of high-grade lenses can be achieved by combining these posteriors, for example, $p(\{A,B\}|\{C_j\})$ rather than $p(A|\{C_j\})$. Notably, the GJ posteriors also have non-negligible scatter, and are slow to converge for intermediate grades. This is indicative of disagreement between researcher graders, showing that with small groups of researchers ($\lesssim5$), the final grade assigned to a system is subject to considerable uncertainty. Obtaining more classifications for systems with an intermediate grade would be useful to reduce this, but it would not solve the underlying issue: grades are merely a rough approximation for lens likelihood and grader confidence. In the future, replacing this grading system with calibrated lens probabilities \citep{Holloway2024} based on a ground-truth data set of confirmed lens/non-lens systems (e.g., using 4SLSLS spectra) would enable large-scale inference using impure but calibrated samples of lenses.

The SWAP pipeline treats each volunteer individually, assigning an individual skill value to each. The uncertainty associated with individual volunteers' skill can be reduced by requiring volunteers to perform a large number of classifications of images with a pre-determined researcher grade and using these classifications for calibration. However, doing so reduces the number of new systems for which volunteer grades can be obtained for the same inspection budget. 
% As an alternative, one can treat all volunteers as a single super-user, with an aggregated skill level calculated in the same manner as previously (Eq.\,\ref{Eq: User Skill}), but where the calibration classifications from all volunteers are used to calculate a single skill matrix rather than one per volunteer. 
Alternatively, all volunteers can be treated as a single super-user using the calibration classifications from all volunteers to produce a single skill matrix (rather than one per volunteer), in the same manner as previously (Eq.\,\ref{Eq: User Skill}). In turn, this reduces the number of calibration images (i.e., systems with a known ground-truth grade) that need to be shown to the volunteers, since all volunteers contribute to the crowd skill. 
Here, we perform crowd-level calibration using 500 volunteer classifications across any subject of a given grade (totalling 2000 classifications) -- these classifications were then removed when calculating the grade posteriors shown in purple in Fig. \ref{Information_Content}. The convergence rate for this crowd-level calibration remains rapid for Grade X systems, reaching \rt{$p(G_i=X|\{C_j\})=0.85$} after only four classifications. For intermediate grades (B/C), the convergence rate is similar when using crowd-level or individual skill, but is slow in both cases. However, for Grade A systems, convergence is much faster when using individual skill. Increasing the number of calibration systems beyond 500 per grade did not substantially improve performance, while decreasing below this number produced slower convergence. Overall, while there are benefits to crowd-level skill (reducing the calibration requirements), analysis using individual skill provides the greatest information for the systems of most interest.

\begin{figure*}
\centering
\includegraphics[width=\textwidth]{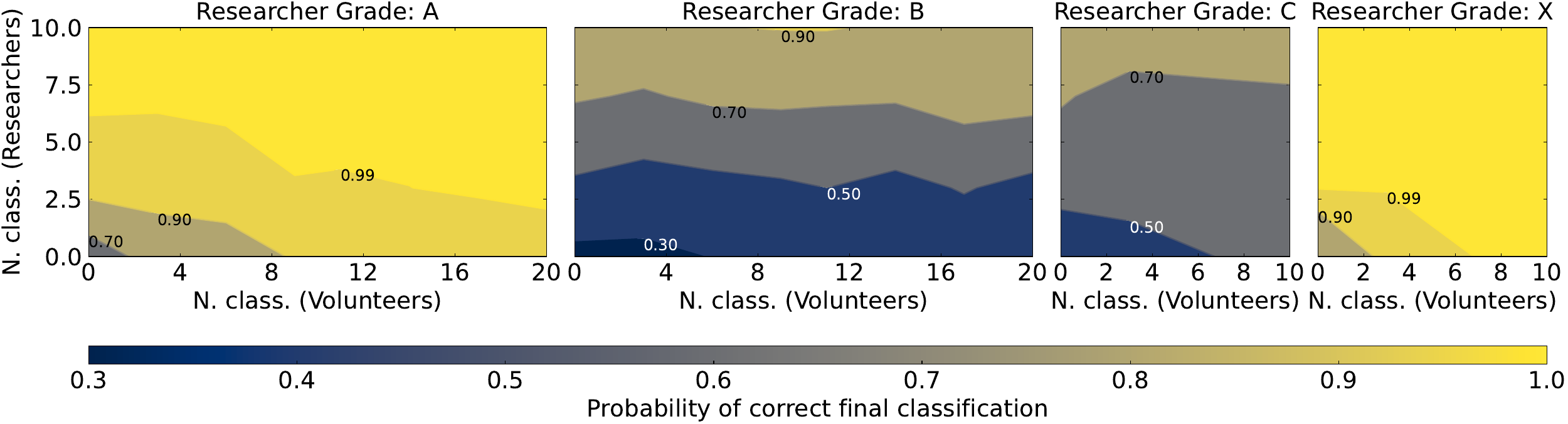}
\caption{Comparison of the information content of \SWRt{} versus GJ classifications. Six classifications from \SWCt{} are included throughout, as this is the pre-cursor stage to both GJ and \SWRt{}. The shading in each panel represents the median grade posterior having received a given number of classifications from \SWRt{} and GJ. For example, the left-hand panel shows the \rt{median} posterior probability $p(G_i=A|\{C_k\}_{\rm{SWC}},\{C_j\}_{\rm{SWR}},\{C_l\}_{\rm{GJ}})$ calculated for systems that received a Grade A consensus from GJ. The posteriors for Grade X systems require the fewest classifications from either GJ or \SWRt{} before convergence.}
\label{Strategies}
\end{figure*}
The convergence of the \SWRt{} grade posteriors in Fig.\,\ref{Information_Content} shows that volunteer grades are a useful proxy for researcher grades. Given that the volunteer crowd is much larger and can thus make more classifications, we investigated how many volunteer classifications are required to achieve a level of confidence in the grade equivalent to that obtained from the researchers.
In Fig. \ref{Strategies} we plot the median grade posterior for systems of a given consensus researcher grade, as a function of the number of classifications received from \SWRt{} and GJ. We include six classifications from \SWCt{} when calculating these posteriors, since such classifications would likely occur prior to \SWRt{} or GJ - as shown by the left-hand panel of Fig. \ref{Conf_Matrix}, even binary classifications carry information as to the grade of a system. To do so we used a new volunteer skill, following Eq.\,(\ref{Eq: User Skill}) but with $C\in\{\rm{Lens,Non}$-$\rm{lens}\}$ rather than $C\in\{\rm{A, B, C, X\}}$ and updated the posterior via Eq.\,(\ref{Eq: Grade_posterior_normalised_by_sum}). In the case of Grade X systems, we find that six classifications from \SWCt{}\footnote{For \SWCt{}, $C_i\in\{\rm{Lens,Non}$-$\rm{lens}\}$.} are already sufficient to give high confidence, with \rt{$p(G=X|\{C_i\})\gtrsim0.7$}, reaching \rt{$p(G=X|\{C_i\})=0.97$} following five classifications from \SWRt{},\footnote{For \SWRt{}, $C_i\in\{\rm{A,B,C,X}\}$.} without any GJ classifications. In such instances, obtaining classifications from researchers via GJ is not merited. Given that such Grade X systems comprised the majority of graded systems in Q1, this could substantially reduce the researcher inspection requirements in future releases.

\subsubsection{Extrapolating to DR1 and the full EWS}\label{S: Extrapolating to DR1/EWS}
\begin{figure*}
\centering
\includegraphics[width=\textwidth]{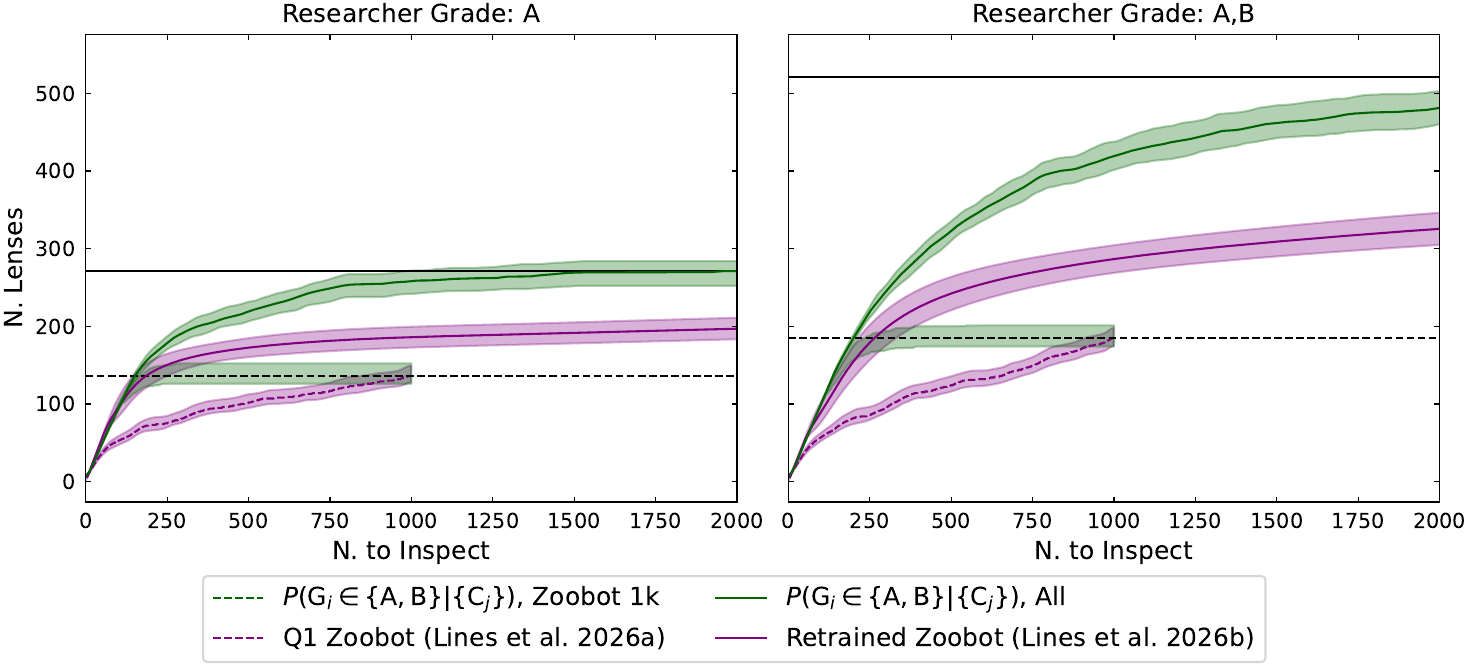}
\caption{
Number of lens candidates (of researcher Grade A and A+B respectively) that would be identified from inspecting the highest-ranked systems from a range of classification methods. Systems ordered by their \SWRt{} grade posterior are plotted in \rt{green}, restricted to the top $1000$ Zoobot candidates (dashed) and compared to the full \SWRt{} sample (solid). The equivalent curves for Zoobot are also shown, using the Q1 Zoobot model (SLDE-C, \rt{purple, dashed}) and a Zoobot model retrained on Q1 lenses (\citealp{Lines2025b}, purple, solid). The confidence regions were determined by bootstrapping and show $\pm1\sigma$ uncertainties. The curve for the retrained Zoobot was evaluated on Q1 data using $k$-fold cross-validation so as to not evaluate performance on the training set; we have subsequently scaled the $y$-axis values for this curve to match the total number of lenses in the \SWRt{} sample.
}
%  \rt{Re-trained Zoobot curve adapted from EC: Lines et al. 2025) } \rt{CITE Lines et al. new paper.}

\label{ROC_Curves}
\end{figure*}
We now consider the benefits of applying \SWRt{} at scale. Figure \ref{ROC_Curves} shows the rate of lens retrieval, as a function of the number of systems inspected, based on the grade posteriors from \SWRt{}. As above (Sect. \ref{Sect: Optimising the inspection strategy}) the volunteers' skill used to determine these posteriors was calculated using the leave-one-out method to improve skill calibration. The values in Fig. \ref{ROC_Curves} are not scaled up to the DR1 area and so represent the performance if the Q1 search (and selection) was rerun. We find that combining the grade posteriors $p(A\cup B|\{C_j\})$ produces the highest purity sample (defined as the ratio of the number of lenses to the total number of systems above a given threshold), likely reflecting the subjective nature of the grading scheme. In this case, \rt{76\% (76\%)} of researcher Grade A (A+B) systems can be retrieved with 50\% purity. This represents a factor of 7 improvement in purity compared to the original GJ sample, in which
% (294+252+20)/(8997+48-1360)
\rt{$7\%$} of the systems were Grade A+B. We note that in the Q1 lens search, the GJ inspection was conducted concurrently with the \SWCt{} inspection. If, instead, the GJ inspection had taken place following the \SWCt{} inspection, the number of systems requiring inspection by GJ could have been reduced, informed by the volunteers' classifications. Approximately 60\% of systems inspected by GJ in Q1 were graded X by researchers and also rejected in \SWCt{} -- these were inspected by GJ due to receiving high ML scores.

Alongside these lens retrieval rates based on grade posteriors, Fig. \ref{ROC_Curves} also shows equivalent curves for two ML models. The first, `Q1 Zoobot', was the best performing network in the Q1 search (referred to as Model 4 in SLDE-C). The second, `Retrained Zoobot', is presented in \citet{Lines2025b}, and uses the lenses identified in Q1 to improve the network's performance. In this case, the performance was measured by excluding subsamples of the Q1 lenses from the network training set. In SLDE-A, the top 1000 systems from each network were inspected by GJ. For the Q1 Zoobot, this corresponded to
%%Removing mention of purity here as it can be ambiguous if I mean cumulative or running purity:
% inspecting down to a Grade A+B purity of approximately $20\%$, 
identifying 
140 %i.e. 136
(190) %i.e. 185
Grade A (A+B) systems, as shown by Fig. \ref{ROC_Curves}. 
% By comparison, the retrained Zoobot would be expected to find 150 and 280 Grade A and Grade A+B lenses respectively within the top 1000 systems (Fig.\,\ref{ROC_Curves}).
Applying \SWRt{} would result in a $85\%$ pure sample retaining $\rt{85\%}$ of the Grade A+B systems in the top 1000 Q1 Zoobot sample. 
% In DR1 and EWS, this would correspond to 
% % 0.85*173*1900/63.1 = 4400
% % 0.85*158*14000/63.1 = 32.6k
% \rt{4400} and \rt{$33\,000$} A+B Grade lenses respectively.
Broadening the selection to the full GJ/\SWRt{} sample, rather than the top 1000 of the Q1 Zoobot, would identify 
420 Grade A+B lenses (solid green, Fig.\,\ref{ROC_Curves}), compared to 
290 retained in the top-scoring 1000 systems from the retrained Zoobot (solid purple). Scaling up the former to the DR1 area (\num{1900}\,$\deg^2$ compared to $63\deg^2$ in Q1) would give \num{12500}
Grade A+B lenses in DR1. This assumes the same pre-selection, lens occurrence rate, and volunteer response behaviour as in Q1. However, we note there is a selection effect here, since \SWRt{} was not applied to the whole Q1 dataset but merely the highest scoring systems from the various ML models in which there are fewer false-positives whereas the Zoobot models were applied to the whole Q1 dataset. However, given such a selection could be employed in the future to larger datasets, such a comparison remains reasonable.
In summary, a citizen-science-based refinement stage can substantially increase the purity of the resultant lens sample. While improvements in ML performance since Q1 have been significant, they do not yet outperform such a refinement step.

As an alternative, \SWRt{} could also be used to remove Grade X systems (definite non-lenses) following \SWCt{}. Setting a threshold of $p(G_i=X|\{C_j\}_{\rm{SWR}})>0.9$ would remove $70\%$ of Grade X systems in the GJ sample, while removing only $1.5\%$ of Grade A+B systems. This would have saved $>1\,\rm{day}$'s worth of cumulative researcher classification time (and $45\,000$ classifications) in the Q1 search\footnote{We used the median GJ classification time of 3 seconds when calculating the potential time saved. However, the GJ classification time had a large tail towards longer times, so this should be seen as a lower limit for the potential GJ inspection time saved.}, which will only increase for forthcoming larger data releases, and would ensure the vast majority of high-grade systems were flagged for researcher inspection.

\subsection{Volunteer selection function}
% - Having applied it twice-over, from SWC and SWR.
\begin{figure*}
\centering
\includegraphics[width=\textwidth]{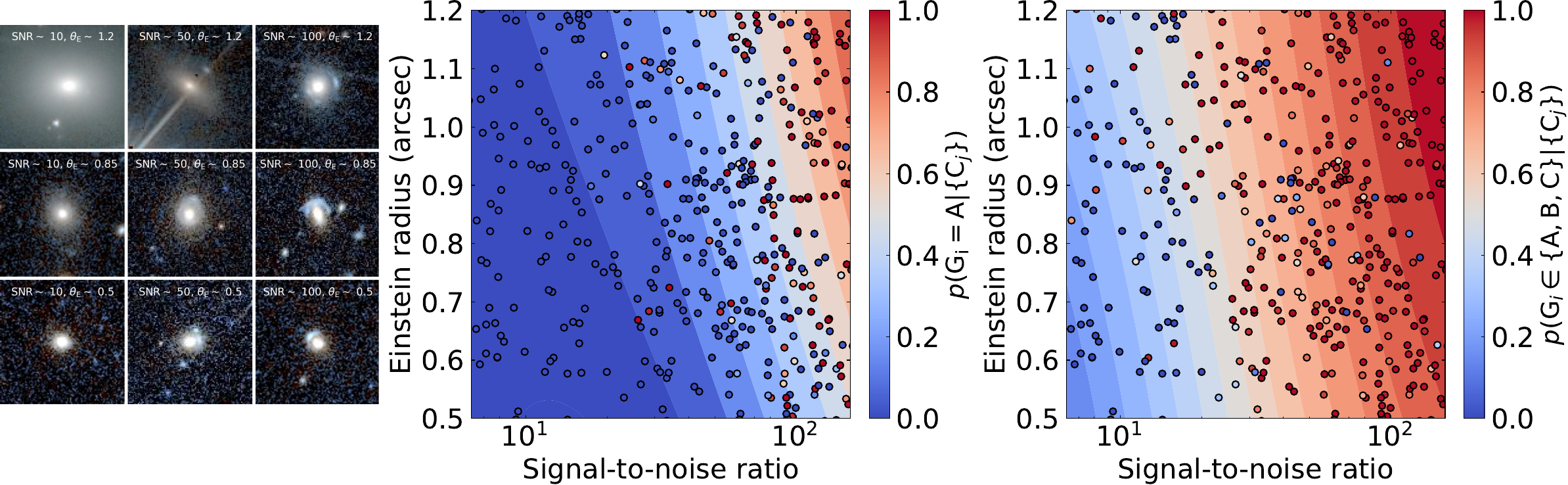}
\caption{ \emph{Left:} Collage of simulated lens systems spanning a range of Einstein radius $\theta_{\rm{E}}$ and arc S/N values. \emph{Centre/Right:} Selection function of the \SWRt{} volunteers showing the probabilities that systems are labelled as Grade A or Grade \rt{A+B+C}, respectively. The points show individual simulated lenses, indicating the grade posterior, while the contours show the smoothed equivalent. Systems with a high S/N are typically assigned a high grade by volunteers (as shown by the red regions), especially if they have a large Einstein radius. Systems with lower S/N can still be identified by volunteers, but are typically assigned a lower grade. }
\label{Selection_Function}
\end{figure*}
The simulations included in the \SWRt{} workflow enable us to test the selection function of the injected simulated lenses: which simulated lens systems the volunteers are likely to classify as lenses, and which they are likely to miss. This is important for studies of the properties of galaxies (see, for example, \citealp{Sonnenfeld2024}) to quantify the differences between lens galaxies (which may be unusually massive or compact), and the typical galaxy population. This selection, as a function of Einstein radius and arc signal-to-noise ratio (S/N) is shown in Fig.\,\ref{Selection_Function}. The S/N was calculated as the maximum value of the ratio of the cumulative lensed source flux and cumulative sum of the root mean square background noise \citep{Collett2015}. 
The primary driver behind whether a system is identified as a lens by volunteers is S/N; systems with a sufficiently high S/N are likely to be identified regardless of their Einstein radius. This is qualitatively similar to the simulated-lens selection function for \SWCt{} (see SLDE-A), which implies that the volunteers do not significantly change their behaviour when performing grading versus binary classification. We find that systems with lower S/N are less likely to be identified as lenses by the volunteers, and those that are identified are likely to be assigned a lower grade. The researchers responded similarly when shown simulated lenses with varying S/N in GJ.
% As shown by Figs. \ref{Conf_Matrix} and \ref{Grade_Bar_Charts}, the responses of the volunteers show some differences to those of the researchers but can be calibrated. 

% \section{Discussion}\label{Discussion}
% \input{5_Discussion}

\section{Conclusions}\label{Conclusion}
We have presented the results from Space Warps Refine, a citizen science project to assign grades of definite, probable, possible, or non-lens to lens candidates from the Euclid Q1 data release. Our conclusions are as follows.
\begin{itemize}
    \item The volunteer crowd responses correlate with those from researchers; systems ranked highly by researchers receive correspondingly high ranks from the volunteers (Fig. \ref{Grade_Bar_Charts}).
    \item Volunteer grades are most useful for identifying Grade A and Grade X systems; the grade posteriors $p(G_i|\{C_j\})$ converge the fastest (i.e., with the fewest volunteer classifications) for these grade extremes (Fig. \ref{Strategies}).
    \item Volunteer grading can provide significant improvements in lens purity. The high-scoring sample from volunteers had \rt{$76\%$ ($76\%$)} completeness of Grade A (A+B) lenses, with a $50\%$ pure sample. This improves upon the original SLDE-A Q1 sample allocated for researcher inspection, of $7\%$ purity of Grade A+B systems and $100\%$ completeness by this paper's ground truth definition. 
    \item The volunteers have a qualitatively similar selection function to the researchers. This holds when both grading and performing binary (lens/non-lens) classification. Systems with low S/N, $\rm{\log_{10}{S/N}}\approx 3$, can still be identified by volunteers but are typically assigned a lower grade (i.e., a less certain lens). 
    \item Binary (lens/non-lens) classification also provides useful information on the lens grade -- higher grade systems received higher scores from \SWCt{}.
\end{itemize}

In summary, volunteer grades can act as a useful proxy to researcher grades. Furthermore, volunteers can inspect significantly larger numbers of systems (of order $10^5$) than researchers alone. Additionally, citizen scientists can reduce the number of classifications required from researchers in order to confidently classify them as Grade A or Grade X lenses, thus reducing the time burden for researchers to perform lens grading which will become more significant with forthcoming \Euclid data releases. 

Euclid DR1 will enable the discovery of thousands of lens candidates. A refinement stage, such as the one presented here, will help maximise the number of lenses found in such a data release, while minimising the inspection burden for both researchers and volunteers.
% \input{6_Conclusion}
% Add the acknowledgement using the achnowledgements environment.
% Do not use \acknowledgement{....} as this affects the formatting
% of the references.
%

\begin{acknowledgements}
We gratefully acknowledge the Zooniverse team for their invaluable work on the host platform for Space Warps. 
This project would not be possible without the immense contributions of citizen scientists. Those contributing classifications for this project are listed in Appendix \ref{Ap: SW Volunteers} and on the Space Warps \href{https://www.zooniverse.org/projects/aprajita/space-warps-esa-euclid/about/team}{`The Team'} webpage. We greatly appreciate their commitment, enthusiasm, and time.
PH and TEC are funded by the Royal Society through a University Research Fellowship.
This work has received funding from the European Research Council (ERC) under the European Union's Horizon 2020 research and innovation programme (LensEra: grant agreement No 945536).

%\AckERO 
\AckQone
\AckDatalabs
\AckEC
\end{acknowledgements}

%
% Here comes the reference list, generated via bibtex from
% your bibfile my.bib and Euclid.bib. Please make sure that
% the same paper is not referenced twice, one from your my.bib
% file, and once from Euclid.bib.
%

\bibliography{References,Euclid}
%
% Now you can add appendices.
% In this example, the appendices are in one column mode.
% If that is not requires, comment out \onecolumn
% Note that appendices in A\&A come {\it after\/} the references.

\begin{appendix}
% \onecolumn %If you don't want single column for the Appendix, please comment this out
\section{Volunteer grades for selected \textit{Euclid} systems}\label{Ap: Cutouts_and_Histograms}
\begin{figure*}
\centering
\includegraphics[width=\textwidth]{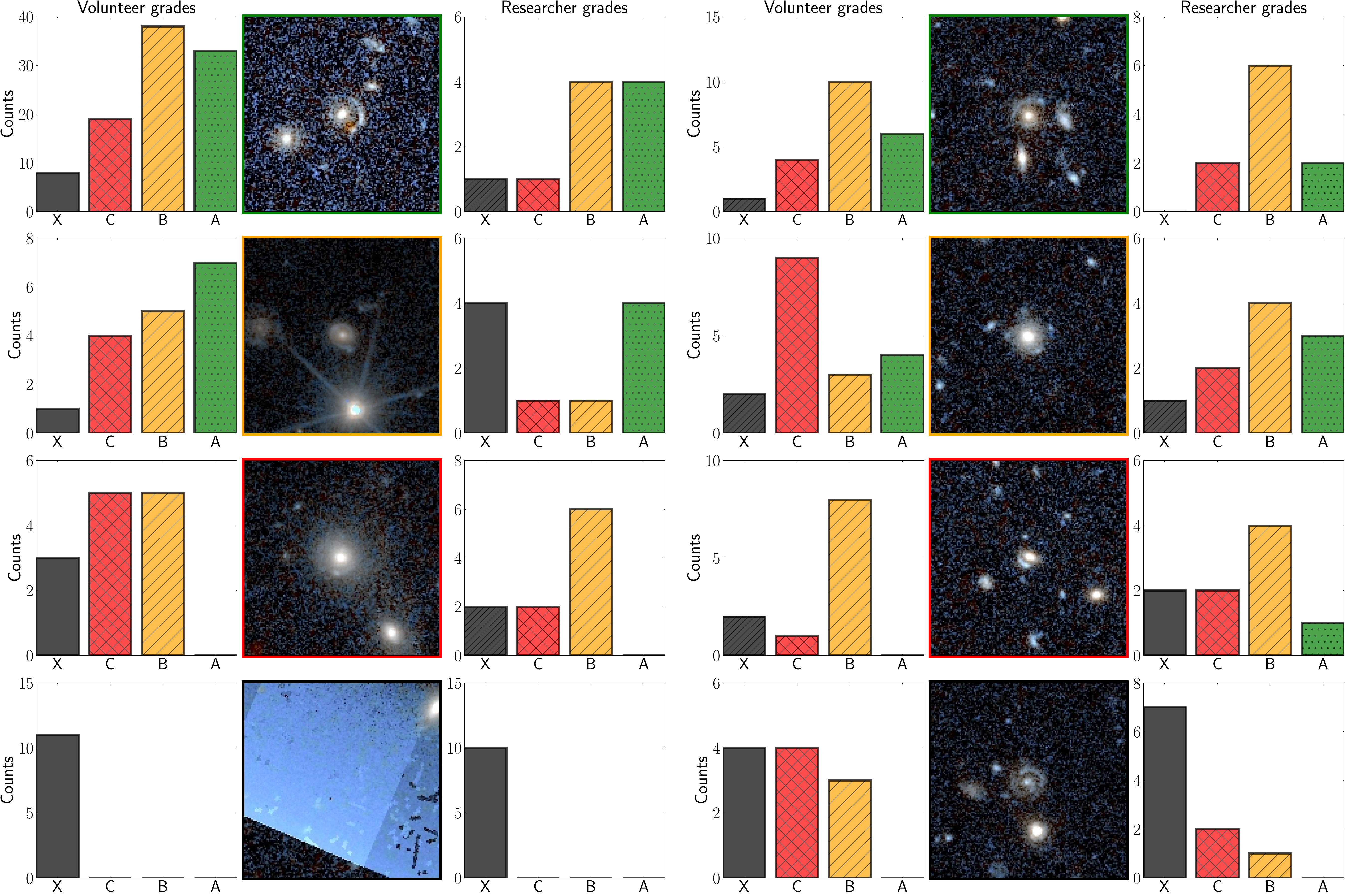}
\caption{Random selection of systems shown to both researcher and volunteer graders, along with the corresponding number of grades received from each. The cutouts are highlighted by their consensus researcher grade: A; green, B; orange, C; red, X; black.}
\label{Random_Grade_Examples}
\end{figure*}
Figure \ref{Random_Grade_Examples} shows a comparison of the distribution of grades received from the researchers and volunteers for a random selection of systems. The cutouts are highlighted by the consensus grade assigned in SLDE-A from GJ. The researchers typically provide more assertive grades (definite lens, or definite non-lens) compared to the volunteers. 

\section{Space Warps volunteers} \label{Ap: SW Volunteers}
This project would not be possible without the immense contributions of citizen-scientists. Those contributing classifications for this project are listed below, and on \href{https://www.zooniverse.org/projects/aprajita/space-warps-esa-euclid/about/team}{`The Team'} webpage. We greatly appreciate their commitment, enthusiasm, and time.
\\
\begin{spacing}{0.7}
\begin{CJK}{UTF8}{gbsn}
\tinier{\noindent Abbey Jennings, Abigail Ginsberg, Ada Rosenfield, Al Lamperti, Alan M. Gross, Alan W Upchurch, Aleksandr Timofeev, Alessandro Bejinaru, Alex Arktos, Alex LaCoss, Alexandre Portefaix, Alexsey Gorbkov, Alissa Strauf, Amanda Peters, Amoli Kakkar, Andreas Thöny, Andrey CRB Gorokhov, André Holland, Angela Karunakaran, Angelin Vincent, Anja Verhöfen, Anna Kozak, Anthony Charest, Antonio Pasqua, Arabella Brosnan, Arkhipova Daria, Arndt Stelter, Ashley McCann, Astro Yona VII, Athene Bunch, Autumn Pearce, Ava Johnston, Bailey Kinner, Barbara Pavlov, Ben Adamowski, Benjamin Gill, Benjamin Graetz, Brian Olivero, Brigitte Wollenweber-Ratzer, Bruno Chiaranti, Bryan Ranharter, Byunghee Kim, C. Graham, Callum Baird, Callum Blake, Camelia Mallek, Caoimhe Murphy, Chaithanya K Iruvaram, Cheng Yılın, Chiriac Eduard Cristian, Chris Molloy, Christopher Doran, Christopher Tanner, Ciro Sirio Perrella, Claire Duguin, Claude Cornen, Colby Jones, Connor Bruce Campbell, Corina Trifu, Cory O'Brien, Coy Morris, Daniel Gascón Jiménez, Daniela Mantula, Danielle Hoss, Danika Belzile, Daria Dawson, David Stegemann, David Truckenmiller, Deepti Sharma, Derek Blauth, Diego Pirrami, Dmitry Ivanov, Donald K. Lorenzo, E Chillington, Eleanor Aalders, Elisabeth Baeten, Elizabeth Beckwith, Emese Kocsis, Emma Gerntke, Eric Fabrigat, Eve England, Evelyne Stanton, Fan Baoyi, Fernanda Freitas, Filip Vdović, Floortje van Santen, Fox James, Frank Falor, G. Greene, Haoxuan Zhang, Harold Price, Heiko Gericke, Helga Haase, Heraldo Henrique Felix de Moraes, Hiago Jorge Silva, Huang Rui Guo, Huub Mooren, Ian Carley, Ingo Holetzek, Ingrid Nurminen, Isabella Ana Cuan, Jaana Kemppainen, Jacob Amato, Jacquelyn Higinbotham, James Lambourne, Jean-Francois Peyron, Jean-Noel Terry, Jennifer Saar, Jens Mario Schmidt, Jessie Huang, Jhoanna Jimena Chanta Flores, Joanna Maria Dulowska, Joe Schmidt, Johnny Reid, Jordi Solaz, Jose Luis Navarro, Joshua McArthur, José Manuel Leal Quesada, Jude Wechsler, Judy Garstone, Juliana Newman, Julien Lühning, Jur Vogelzang, Justine Lefebvre, Jérôme F. Delaunay, Kai Schindler, Karthikeyan Duraisamy, Katharina Stowe Martin, Katrina Wong, Ken Davis, Ken O'Neill, Kevin Peter, Kostula Papandrianu, Ksenia Smirnova, Lamberto Veraldi, Lauren Li, Lauren Murphy, Laëtitia Renard, Leishangthem Debajit Singh, Liam Buki, Lianne Louise Manalili, Liew En Quan, Lino Thomas, Louisa Kupke, Loïc Baert, Lysandros Wasilopoulos, M. Anderson, Maggie Jardine, Mai Tanaka, Maia G. Mohsen, Marc Hanset, Marco Bacchi, Maria Eduarda Alves de Araújo, Maria Tubol, Mariella Gozzi, Martina Cucchi, Matayah G, Matthew Allis, Matthew Loomis, Michael Mulhern, Michele T. Mazzucato, Molly Scott, Monica Backes, Monika Seniut, Muhammad Muzammil Anwar, Natalie Freund, Natalie Savelyeva, Nathan Corey Bell, Nerea Ruiz Martín, Nicola Denti, Nikolas Kontaxakis, Noah R, Nolan Reket, Olivia Park, P Gopalan, P. Seal, Paola Bungaro, Pedro Bergevoet, Prasad Vengurlekar, Preetha Pushkarini B N, Quentin Guelle, Rachel Alice Ford, Rainer Gorissen, Rainer Töpler, Rebecca Denae, Rebecca Eid, Retaj Atef Elsobky, Richard Goodwin, Richard Knijnenburg, Rishaan Akula, Robert J. Adamson, Roland Heringhaus, S. Ominami, Samanta Holdaway, Sana Farhat Abdul Qayyum, Sandra Lee Harris, Sanjit Kolluru, Sara I. Estrada, Sarah Grissett, Sarp Akar, Satanay Makhmud, Sauro Gaudenzi, Shneor Gubits, Silke Wachholz, Simon Nicholson, Simone Glau, Soveros Arctos, Stacey Cornell, Stacey Cornell, Stefanie Seidel, Stefanos Drakoulis, Steve Saunders, Susannah Stubbs, Sushant Pandit, Svetoslav Alexandrov, Sylvain Martineau, Tanja Kuusela, Thomas Rosa, Thorsten Eschweiler, Titiksha Pandit, Vansh Bahety, Victor Juan Garcia Porcel, Vincent Parmentier, Virgilio Gonano, Vishakha Jingar, Wolfgang G. Kurtz, Yang Feichi, Zoe Posnette, \foreignlanguage{greek}{ΒΑΣΙΛΙΚΗ ΠΕΤΡΟΠΟΥΛΟΥ}, \foreignlanguage{russian}{Цыро Снежана}, 片山洋臣 Yoshifumi Katayama}
\end{CJK}
\end{spacing}
\end{appendix}

\end{document}